\PassOptionsToPackage{table,dvipsnames,xcdraw}{xcolor} 

\documentclass[a4paper]{article}

\usepackage[english]{babel}
\usepackage[utf8x]{inputenc}
\usepackage[T1]{fontenc}

\usepackage[a4paper,top=3cm,bottom=2cm,left=3cm,right=3cm,marginparwidth=2.00cm]{geometry}

\usepackage{amsmath}
\usepackage{amssymb}
\usepackage{amsfonts} 
\usepackage{bm} 
\usepackage{graphicx}
\usepackage{xcolor}
\usepackage[colorlinks=true, allcolors=blue]{hyperref}

\usepackage{xurl} 

\usepackage{parskip} 
\usepackage{cleveref} 
\usepackage{enumitem} 
\usepackage{multirow}
\usepackage{adjustbox} 

\usepackage[superscript,biblabel]{cite} 

\DeclareMathOperator{\EX}{\mathbb{E}} 

\newcommand{\R}{\mathbb{R}} 

\usepackage{graphicx}
\newsavebox\CBox

\usepackage{float} 

\usepackage{authblk} 

\title{Evaluating individualized treatment effect predictions: \protect\\a model-based perspective on discrimination and \protect\\calibration assessment}

\author[1,2]{J Hoogland}
\author[3,4]{O Efthimiou}
\author[5]{TL Nguyen}
\author[1,7]{TPA Debray}

\affil[1]{Julius Center for health sciences and primary care, University Medical Center Utrecht, Utrecht University, Utrecht, the Netherlands}
\affil[2]{Epidemiology and Data Science, Amsterdam University Medical Center, Amsterdam, the Netherlands}
\affil[3]{Institute of Primary Health Care (BIHAM), University of Bern, Bern, Switzerland}
\affil[4]{Institute of Social and Preventive Medicine (ISPM), University of Bern, Bern, Switzerland}
\affil[5]{Section of Epidemiology, Department of Public Health, University of Copenhagen, Copenhagen, Denmark}
\affil[6]{Cochrane Netherlands, University Medical Center Utrecht, Utrecht University, Utrecht, the Netherlands}
\affil[7]{Smart Data Analysis and Statistics B.V., Utrecht, the Netherlands}

\date{}                    

\begin{document}


\maketitle

\begin{abstract}
In recent years, there has been a growing interest in the prediction of individualized treatment effects. While there is a rapidly growing literature on the development of such models,  there is little literature on the evaluation of their performance. In this paper, we aim to facilitate the validation of prediction models for individualized treatment effects. The estimands of interest are defined as based on the potential outcomes framework, which facilitates a comparison of existing and novel measures. In particular, we examine existing measures of measures of discrimination for benefit (variations of the c-for-benefit), and propose model-based extensions to the treatment effect setting for discrimination and calibration metrics that have a strong basis in outcome risk prediction. The main focus is on randomized trial data with binary endpoints and on models that provide individualized treatment effect predictions and potential outcome predictions. We use simulated data to provide insight into the characteristics of the examined discrimination and calibration statistics under consideration, and further illustrate all methods in a trial of acute ischemic stroke treatment. The results show that the proposed model-based statistics had the best characteristics in terms of bias and accuracy. While resampling methods adjusted for the optimism of performance estimates in the development data, they had a high variance across replications that limited their accuracy. Therefore, individualized treatment effect models are best validated in independent data. To aid implementation, a software implementation of the proposed methods was made available in \texttt{R}.

\end{abstract}


\section{Introduction}

Prediction models for important health outcomes have long been a crucial aspect of personalized healthcare \cite{harrell_multivariable_1996, steyerberg_clinical_2019, harrell_regression_2015}.  In line, methods for assessing their performance have been well established, and include overall accuracy, discrimination, and calibration assessment \cite{harrell_multivariable_1996, steyerberg_clinical_2019, harrell_regression_2015, van_calster_calibration_2016, van_calster_calibration_2019}.

In recent years, there has been a growing interest in the prediction of key health outcomes under different treatment options \cite{kent_personalized_2018, kent_predictive_2020, senn_statistical_2018, rekkas_predictive_2020, lin_scoping_2021, hoogland_tutorial_2021}. Such individualized treatment effect (ITE) predictions are clearly of interest in many applied settings, including medical decision making. Due to the causal nature of ITE predictions, such models are typically developed in randomized data or explicitly account for confounding by other means. While there is a large and rapidly growing literature on the \textit{development} of models for individualized treatment effect prediction, $^{\textit{e.g.}}$ \cite{lin_scoping_2021, hoogland_tutorial_2021, tian_simple_2014, chernozhukov_generic_2018, wager_estimation_2018, athey_generalized_2019, kunzel_metalearners_2019, nie_quasi-oracle_2021} literature on the corresponding \textit{performance assessment} is scarce \cite{hoogland_tutorial_2021, van_klaveren_proposed_2018, efthimiou_measuring_2023, maas_performance_2023}. 

In this paper, we build on existing proposals for the assessment of clinical prediction models for ITE prediction \cite{van_klaveren_proposed_2018, efthimiou_measuring_2023, van_klaveren_measuring_2023}, with a focus on measures of discrimination for benefit and calibration for benefit. Discrimination and calibration have been key quantities of interest for clinical prediction model evaluation in the case of binary outcomes \cite{harrell_multivariable_1996, steyerberg_clinical_2019, van_calster_calibration_2016, harrell_evaluating_1982}, and previous proposals for discrimination for benefit and calibration for benefit have mainly focused on binary outcomes \cite{van_klaveren_proposed_2018, van_klaveren_measuring_2023}. Consequently, we focus on binary outcomes, which are also very common in randomized clinical trials \cite{charles_reporting_2009}. Nevertheless, we will briefly digress into analogous procedures for continuous outcomes, since both quantities are also of interest there as well. 

With respect to the types of ITE prediction models of interest, we focus on causal prediction models that contrast outcome predictions under different treatment options. This in contrast to (i) models that directly estimate ITEs without considering the outcomes per treatment condition as key estimands, and (ii) models that only predict the sign of treatment effect (\textit{i.e.} benefit or harm). The main reason is that medical decision making is best informed by \textit{both} prognostic information \textit{and} treatment effect information, and should balance the benefits and harms of initiating particular treatments, the underlying risk of disease without treatment, and patient preferences \cite{sacket_evidence_1996}. For example, an effective treatment with significant potential harm and cost may be of interest only to those at high risk without the treatment \cite{alba_discrimination_2017}. 

Previous work by our group has addressed a number of evaluation metrics for ITE models for both binary and continuous outcomes, including decision accuracy, discrimination for benefit and calibration for benefit \cite{efthimiou_measuring_2023}. The main contribution of this paper is a more in depth exploration of discrimination and calibration for benefit that (i) utilizes the potential outcomes framework to increase clarity of exposition and to clearly define the estimands of interest \cite{rubin_estimating_1974, rubin_causal_2005}, (ii) incorporates recently proposed modifications to the c-for-benefit \cite{van_klaveren_measuring_2023, maas_performance_2023}, and (iii) proposes model-based estimators of the defined discrimination and calibration estimands that avoid the need for matching. Simulation results are provided for illustrative purposes. An applied example using data from the third International Stroke Trial (IST-3) \cite{the_ist-3_collaborative_group_benefits_2012} serves to further illustrate implementation in practice. To aid implementation, a software implementation of the proposed methods is made available in \texttt{R}.

\section{Individualized treatment effect prediction} \label{sec:ite}

Most outcome prediction research focuses on capturing statistical association in absence of interventions. Individualized treatment effect (ITE) prediction is a different type of prediction since it has a causal interpretation: the quantity to be predicted is the effect caused by the treatment (or intervention, in a larger sense) on the outcome. Therefore, before moving to the performance measures of interest, this section shortly outlines causal prediction. Subsequently, issues surrounding the use of binomial outcome data for ITE modeling are shortly discussed (further details are available as online supplementary material \ref{app:binom_challenges}).

\subsection{Causal prediction} \label{sec:causal_prediction}

To emphasize the causal nature of the predictions, it is helpful to write the individualized treatment effect of interest in terms of the potential outcomes framework \cite{rubin_estimating_1974, rubin_causal_2005}. For treatment taking values $a \in \mathcal{A}$, $Y^{A=a}$ denotes the potential outcome under treatment $a$. When comparing two treatments, the ITE for individual $i,\ldots,n$ can be defined as 
\begin{equation} \label{eq:deltai_potential}
    \delta(\bm{x}_i) = \EX(Y^{a=1} | \bm{X} = \bm{x}_i) - \EX(Y^{a=0} | \bm{X}_i = \bm{x}_i)
\end{equation}
where $\bm{x}_i$ is a row vector of individual-level characteristics in matrix $\bm{X}$. The degree of granularity or individualization reflected by $\delta(\bm{x}_i)$ relates to the number of predictors included in $\bm{X}$, to the strength and shape of their association with the potential outcomes, and especially to the degree to which they have a differential effect across potential outcomes (\textit{i.e.}, modify the effect of treatment). Ideally, the set of measured individual-level characteristics includes all relevant characteristics with respect to individualized treatment effect. In practice however, this set of all relevant characteristics is often unknown and the best way forward is to aim for conditioning on the most important characteristics. Correspondingly, equation \eqref{eq:deltai_potential} reflects ITE as a conditional treatment effect for some set of characteristics. 

Since in practice only one potential outcome is observed per individual \cite{holland_statistics_1986}, assumptions are required to estimate $\delta(\bm{x}_i)$ based on the observed data. These assumptions are discussed in detail elsewhere \cite{hoogland_tutorial_2021, hernan_causal_2020}. In short, the key assumptions are \textit{exchangeability} (the potential outcomes do not depend on the assigned treatment), \textit{consistency} (the observed outcome under treatment $a \in \mathcal{A}$ corresponds to the potential outcomes $Y^{A=a}$), and positivity (each individual has a non-zero probability of each treatment assignment. An additional assumption that eases inference is \textit{no interference} (the potential outcomes for individual $i$ do not depend on treatment assignment to other individuals). Based on these assumptions, the individualized treatment effect can be identified given the observed data:
\begin{align} 
\delta(\bm{x}_i) &= \EX(Y^{a=1}|\bm{X}=\bm{x_i}) - \EX(Y^{a=0}|\bm{X}=\bm{x_i}) \nonumber \\ 
                 &= \EX(Y^{a=1}| A=1, \bm{X}=\bm{x_i}) - \EX(Y^{a=0}| A=0, \bm{X}=\bm{x_i}) \quad \textnormal{(by exchangeability)} \nonumber \\
                 &= \EX(Y_i| A=1, \bm{X}=\bm{x_i}) - \EX(Y_i| A=0, \bm{X}=\bm{x_i})  \quad \textnormal{(by consistency)} \label{eq:delta_iobs}
\end{align}

Equation \eqref{eq:delta_iobs} shows that ITE predictions $\hat{\delta}(\bm{x}_i)$ can be estimated using a prediction model for outcome risk $\EX(Y_i | A=a_i, \bm{X}=\bm{x}_i)$. Many modeling tools can be used for this endeavor and the details are beyond the scope of this paper and are given elsewhere (\textit{e.g.,} \cite{hoogland_tutorial_2021, lamont_identification_2018}). When conditioning on $\bm{x}$ is clear from the context, we at times abbreviate $\hat{\delta}(\bm{x}_i)$ as $\hat{\delta}_i$, or write the vector of predictions for all individuals $1,\ldots,n$ as $\hat{\bm{\delta}}$.

\subsection{Observed outcome data}
For binary outcomes, we observe outcome $Y_i \in \{0,1\}$ and covariate status $\bm{x}_i$ for each individual $i$. In this context, the ITE estimate $\delta(\bm{x}_i)$ is a difference between two risk predictions ($P(Y_i = 1|A = 1, \bm{X}=\bm{x}_i) - P(Y_i = 1|A = 0, \bm{X}=\bm{x}_i)$). While other quantities are available to express treatment effect (\textit{e.g.}, relative risk, odds ratio), risk differences are generally preferred and better understood by clinicians \cite{bobbio_completeness_1994, naylor_measured_1992, sorensen_laypersons_2008}. The range of $\hat{\delta}(\bm{x}_i)$ includes all values in the $[-1,1]$ interval, while the observed difference between any two outcomes can only be one of $\{-1, 0, 1\}$. Therefore, the observations come with large and irreducible binomial error and provide only limited information. Also, predictions for binary outcome data commonly involve non-linear functions of the covariates, and hence the effects of treatment and the covariates are typically not additive on the risk difference scale of interest here. Consequently, the resulting ITE predictions conflate variability from different sources: between-subject variability in $P(Y^{a=0}=1|X=x)$ and genuine treatment effect heterogeneity on the scale used for modeling. This is the price to pay for the benefit in terms of interpretation of measures on the scale of $\delta(x)$ \cite{murray_patients_2018}. 

In case of a continuous outcome, the problems at hand simplify considerably. First, continuous outcomes are far less noisy than binary outcomes. Second, many continuous outcome models have an identity link function, which puts the parameter effects directly on the outcome scale, and, most importantly, avoids the conflation of the effects of treatment effect related parameter and other model parameters. 

Regardless of the type of outcome measure, the fact that only one potential outcome can be observed is a key challenge at the time of model development \textit{and} evaluation. As opposed to the evaluation of regular prediction models of directly observable outcomes, a direct comparison between predictions $\hat{\delta}(\bm{x}_i)$ and observed outcomes is not feasible.

\section{Discrimination for individualized treatment effects} \label{sec:discrimination}

Discriminative model performance reflects the degree to which model predictions are correctly rank-ordered and is a common performance measure in prediction models for binary and survival endpoints \cite{steyerberg_clinical_2019, harrell_regression_2015}. In the binary endpoint setting, individualized treatment effects estimates $\hat{\delta}(\bm{x}_i)$ for $i,\ldots,n$ range from $[-1,1]$, and differences between potential outcomes $\Delta_i = Y^{a=1}_i - Y^{a=0}_i$ take values in $\{-1, 0, 1\}$. The aim is to quantify the degree to which $\hat{\delta}(\bm{x})$ correctly rank-orders the probability to observe benefit in terms of $\Delta$. The continuous case is discussed in supplementary material \ref{app:continuous}.

\subsection{Discrimination estimand}
We base our estimand on the well-known c-statistic, which describes the proportion of concordant pairs amongst pairs not tied on the outcome. Analogously, for a randomly sampled pair of cases with discordant outcomes, it describes the probability of concordant predictions (\textit{i.e.} for current purposes: where the case with the lower predicted treatment effect in the pair also has the lower probability of benefit). In particular, we start from the c-statistics as formulated by Harrell \cite{harrell_evaluating_1982}, since it allows for ordinal outcomes. Note that such measures of concordance with respect to rank order have a long history, dating back to Kendall's proposal of $\tau$ as a measure of rank correlation. Supplementary material \ref{app:history} provides a short overview of measures of association leading up to the c-statistic by Harrell to build some more intuition. Adapting to the current setting, for a randomly selected pair of cases $k,l \in 1,\ldots,n$ ($k \neq l$) 
\begin{align} \label{eq:conc}
P(\textnormal{concordance } | \textnormal{ differential benefit})
    &= \frac{P(\textnormal{concordance } \cap \textnormal{ differential benefit})}
    {P(\textnormal{differential benefit})} \nonumber \\
    &= \frac{
    P(\Delta_k < \Delta_l \cap \hat{\delta}_k < \hat{\delta}_l) + 
    P(\Delta_k > \Delta_l \cap \hat{\delta}_k > \hat{\delta}_l)}{
    P(\Delta_k < \Delta_l \cup \Delta_k > \Delta_l)}
\end{align}

When the proportion of concordant pairs increases, equation \eqref{eq:conc} goes to 1; conversely, it goes to 0 when the proportion of discordant pairs increases. Likewise, equation \eqref{eq:conc} moves to 0 when the proportion of ties in $\hat{\delta}$ for pairs with a nonzero $P(\Delta_k \neq \Delta_l)$ increases. In line with the c-statistic, we opt for an estimand that moves toward 0.5 for pairs tied on $\hat{\delta}$ and with nonzero $P(\Delta_k \neq \Delta_l)$. In line, our main discrimination estimand of interest can be formulated as 
\begin{align} \label{eq:theta_d}
\theta_d &= \frac{
    P(\Delta_k < \Delta_l \cap \hat{\delta}_k < \hat{\delta}_l) + 
    P(\Delta_k > \Delta_l \cap \hat{\delta}_k > \hat{\delta}_l) + 
    \zeta}{
    P(\Delta_k < \Delta_l \cup \Delta_k > \Delta_l)},
\end{align}

where $\zeta = \frac{1}{2} P(\Delta_k \neq \Delta_l \cap \hat{\delta}_k = \hat{\delta}_l)$. Consequently, 0.5 is represents a neutral value for the case where the order of $\hat{\delta}_k, \hat{\delta}_l$ does not provide information on the probability of benefit (or harm). Since the comparisons are made between all $k$ and $l$ where $k \neq l$, each pair is evaluated twice and benefit in the comparison $k$ to $l$ means harm in the comparison $l$ to $k$. This allows for an equivalent formulation of $\theta_d$ in terms of just benefit as 
\begin{align} \label{eq:theta_d2}
\theta_d &= \frac{
    P(\Delta_k < \Delta_l \cap \hat{\delta}_k < \hat{\delta}_l) + 
    \zeta^*}{
    P(\Delta_k < \Delta_l)},
\end{align}
where $\zeta^* = \frac{1}{2} P(\Delta_k < \Delta_l \cap \hat{\delta}_k = \hat{\delta}_l)$.

\subsection{Discrimination estimators based on matching}
In practice, differences between the potential outcomes of interest are not observable at the individual level. Consequently, the required $\Delta$'s in $\theta_d$ are unavailable and have to be approximated based on the data. Thus, the truly individual $\Delta$'s are out of reach and in the following we will use approximations conditional on covariates $\bm{x}$. One of the possibilities is to use matching as in the proposed c-for-benefit that aims to quantify discriminative performance on the ITE level \cite{van_klaveren_proposed_2018}. Below we outline the c-for-benefit, its properties, and a modification. Thereafter, we introduce an alternative model-based approach to estimate $\theta_d$.

\subsubsection{C-for-benefit definition} \label{sec:c-for-benefit_overview}
In the setting of a randomized two-arm study measuring a binary outcome of interest, the c-for-benefit aims to assess discrimination at the level of ITE predictions (referred to as 'predicted [treatment] benefit' in the original paper).\footnote{The original paper did not focus on the required conditions for \textit{causal} interpretation of the predicted individualized treatment effects; here we assume that these assumptions, as described in Section \ref{sec:causal_prediction}, are met.} The problem of unobserved individual treatment effects is approached from a matching perspective. One-to-one matching is used to match treated individuals to control individuals based on their predicted treatment effects. The subsequent data pairs hence consist of a treated individual and a control individual with similar predicted treatment effect. Observed treatment effect within the pair is defined as the difference in outcomes between these two individuals. Of note, observed (within-pair) treatment effect can only be in \{-1,0,1\}. Subsequently, the c-for-benefit has been defined as "the proportion of all possible pairs of matched individual pairs with unequal observed benefit in which the individual pair receiving greater treatment benefit was predicted to do so" \cite{van_klaveren_proposed_2018}. The predicted treatment effect within each pair used in this definition is taken to be the (within-pair) average of predicted treatment effects. That is, for a control individual $i$ out of $1,\ldots,n_i$ (with $n_i$ the number of controls) and a treated individual $j$ out of $1,\ldots,n_j$ (with $n_j$ the number treated), predicted treatment effects are taken to be

\begin{align} \label{eq:deltahat_ij}
    \hat{\delta}_{ij}(\bm{x}_i, \bm{x}_j) = \{ &(\hat{P}(Y_i =1 | A_i=1, \bm{X} = \bm{x}_i) - \hat{P}(Y_i =1| A_i=0, \bm{X} = \bm{x}_i)) + \nonumber \\ 
    &(\hat{P}(Y_j =1 | A_j=1, \bm{X} = \bm{x}_j) - \hat{P}(Y_j =1 | A_j=0, \bm{X} = \bm{x}_j)) \} /2
\end{align}

The 'observed' treatment effect is subsequently taken to be $O_{ij}=Y_i - Y_j$. The c-for-benefit is an application of the c-statistic by Harrell \cite{harrell_evaluating_1982} as applied to predictions $\hat{\delta}_{ij}(\bm{x}_i, \bm{x}_j)$ and observations $O_{ij}$ from the matched pairs (further details on the derivation of Harrell's c are given in the Supplementary material \ref{app:history}). If the two (binary) outcomes in such a pair are discordant, then there supposedly is some evidence of a treatment effect (\textit{i.e.}, benefit or harm); conversely, there is no such evidence when the outcomes are concordant (\textit{i.e.}, the predicted treatment effect did not manifest as a difference in outcomes). The implicit assumption is that individual $i$ and $j$ are similar enough to serve as pseudo-observations of the unobserved potential outcomes. 

\subsubsection{C-for-benefit properties} \label{sec:c-for-benefit_procedure}
Although the c-for-benefit has been applied on several occasions (\textit{e.g.}, \cite{bress_patient_2021,olsen_which_2021, duan_clinical_2019}), its properties have not been fully elucidated. Van Klaveren et al. \cite{van_klaveren_proposed_2018} recommended further work on its theoretical basis and simulation studies, which we present here. In parallel with our work, Xia et al. have considered related and complementary methodological aspects of the c-for-benefit, which we will also relate to here.\cite{xia_methodological_2023}.

While the c-for-benefit was not developed with our estimand $\theta_d$ in mind (in fact, no estimand was specified \cite{van_klaveren_proposed_2018}), it is helpful to further dissect what it is estimating. As described, the c-for-benefit compares concordance between differences in i) average predicted treatment effect within matched control-treated pairs $\hat{\delta}_{ij}(\bm{x}_i, \bm{x}_j)$ and ii) observed outcome differences within those same pairs $O_{ij}$. In general, however, $\delta_{ij}(\bm{x}_i, \bm{x}_j) \neq \EX(O_{ij}|\bm{x}_i, \bm{x}_j)$ unless $\bm{x}_i = \bm{x}_j$. Specifically, for controls $i \in 1,\ldots,n_i$ and treated individuals $j \in 1,\ldots,n_j$ and writing ${g}_0(\bm{x})$ for $P(Y_i=1| A=0, \bm{X}=\bm{x})$ and ${g}_1(\bm{x})$ for $P(Y_i=1| A=1, \bm{X}=\bm{x}$), 
\begin{align} 
\EX(O_{ij}|\bm{x}_i, \bm{x}_j) =& \EX(Y_j|\bm{x}_j) - \EX(Y_i|\bm{x}_i)  \nonumber \\ 
 =& {g}_1(\bm{x}_j) - {g}_0(\bm{x}_i) \label{eq:EXOij_0} \\ 
 =& [g_0(\bm{x}_j) + \delta(\bm{x}_j)] - g_0(\bm{x}_i) \label{eq:EXOij_a} \\ 
 =& g_1(\bm{x}_j) - [g_1(\bm{x}_i) - \delta(\bm{x}_i)] \label{eq:EXOij_b}
\end{align}

\textbf{Perfect matching} In case of perfect matching, (\textit{i.e.}, $\bm{x}_i = \bm{x}_j$), it can be seen from equations \eqref{eq:deltahat_ij}, \eqref{eq:EXOij_a}, and \eqref{eq:EXOij_b} that $\EX(O_{ij}|\bm{x}_i, \bm{x}_j) =  \delta(\bm{x}_j) =  \delta(\bm{x}_i)$ and $\hat{\delta}_{ij}(\bm{x}_i, \bm{x}_j) = \hat{\delta}(\bm{x}_j) = \hat{\delta}(\bm{x}_i)$. Relating this to our estimand $\theta_d$, perfect matching on $\bm{x}$ provides the required information on $\Delta$ (since $\EX(O_{ij}|\bm{x}_i, \bm{x}_j) = \EX(\Delta_i | \bm{x}) = \EX(\Delta_j |\bm{x})$) and uses the correct treatment effect estimates $\delta(\bm{x}_i)$). Thus, the c-for-benefit based on predictions $\hat{\delta}_{ij}(\bm{x}_i, \bm{x}_j)$ and outcomes $O_{ij}$ estimates $\theta_d$ under perfect matching.

\textbf{Imperfect matching} Two matching procedures were proposed for the c-for-benefit: i) based on $\hat{\delta}(\bm{x})$ (\textit{i.e.}, minimize the distance between pairs $\hat{\delta}_i$ and $\hat{\delta}_j$), and an alternative ii) based on the Mahalanobis distance between covariate vectors \footnote{ \label{footnote:Mahalanobis} where the distance between $\bm{x}_i$ and $\bm{x}_j$ is defined as $d(\bm{x}_i,\bm{x}_j) = \sqrt{(\bm{x}_i-\bm{x}_j)' \bm{S}^{-1} (\bm{x}_i-\bm{x}_j)}$ with $\bm{S}$ the covariance matrix of the covariates $\bm{x}$} \cite{van_klaveren_proposed_2018}. In theory, matching on covariates $\bm{x}$ leads to appropriate matches as described above. However, it is notoriously difficult in case of increasing dimension of $\bm{x}$ and requires appropriate scaling or weighting (importance assignment) for all elements of $\bm{x}$. Matching on $\hat{\delta}(\bm{x})$ is one-dimensional and hence much easier, but does not necessarily lead to appropriate matches on $\bm{x}$ since $\hat{\bm{\delta}}(\bm{x})$ is generally not an injective function of $\bm{x}$ (\textit{i.e.}, multiple configurations of $\bm{x}$ can give rise to the same value of $\hat{\delta}(\bm{x})$). In general, when matches are only approximate in terms of $\bm{x}$, $\EX(O_{ij}|\bm{x}_i, \bm{x}_j)$ is not equal to either $\delta(\bm{x}_j)$ or $\delta(\bm{x}_i)$. Specifically, as most easily seen in equation \eqref{eq:EXOij_a}, $O_{ij}$ will reflect treatment effect in the treated $\delta(\bm{x}_j)$ \textit{and} differences in risk under the control condition between case $i$ and $j$ (\textit{i.e.}, $g_0(\bm{x}_j)$ and $g_0(\bm{x}_i)$ may differ when $\bm{x}_i \neq \bm{x}_j$). Hence, $\EX(O_{ij}|\bm{x}_i, \bm{x}_j) = \Delta(\bm{x}_j) + \xi_{ij}$, where $\xi_{ij} = g_0(\bm{x}_j) - g_0(\bm{x}_i)$ is not related to treatment but to variability in control outcome risk, and will typically have $\EX(\xi_{ij}) \neq 0$. Also, when $\bm{x}_i \neq \bm{x}_j$ leads to $\hat{\delta}(\bm{x}_j) \neq \hat{\delta}(\bm{x}_i)$, $\hat{\delta}_{ij}(\bm{x}_i, \bm{x}_j)$ is no longer equal to either $\hat{\delta}(\bm{x}_j)$ or $\hat{\delta}(\bm{x}_i)$ (but to their average). In terms of our estimand $\theta_d$, (i) the approximation of $P(\Delta_k < \Delta_l$) and $P(\Delta_k > \Delta_l)$ is \textit{too} variable due to $\xi_{ij}$, and (ii) the pairwise averaged $\hat{\delta}_{ij}(\bm{x}_i, \bm{x}_j)$ is \textit{less} variable than the individual level treatment effect estimates that are to be evaluated. While the effect that this may have on rank-ordering is not straightforward, it might at least be expected that the presence of $\xi_{ij}$, that is unexplained by the treatment effects under evaluation, leads to a bias in $\hat{\theta}_d$ toward the neutral value $0.5$.

\textbf{Loss of data} Both matching procedures were proposed for $1:1$ matching, which requires either equal groups size for both study arms or loss of data. A simple remedy that stays close to the original idea is to perform repeated analysis with random sub-samples of the larger arm \cite{efthimiou_measuring_2023}. Alternatively, many-to-one matching (\textit{e.g.}, full matching) or many-to-many matching \cite{rosenbaum_characterization_1991, hansen_full_2004, colannino_efficient_2007} might be implemented, but none of these has been studied in the context of the c-for-benefit. 

\subsubsection{C-for-benefit modifications} \label{sec:cben_mods} 
\textbf{Matching on predicted control outcome risk}
From equation \eqref{eq:EXOij_a} it can be seen that adjusting the pairwise outcome difference $O_{ij}$ based on known $g_0(\cdot)$ leaves just $\delta(\bm{x}_j)$ (in expectation). That is, $O_{ij}^* = (Y_j - g_0(\bm{x}_j)) - (Y_i - g_0(\bm{x}_i))$ has a useful expectation that equals the true individualized treatment effect for the treated individual $j$
\begin{align} \label{eq:delta_trt}
\EX(O_{ij}^* | g_0(\bm{x}_i), g_0(\bm{x}_j)) = \EX(Y_j - g_0(\bm{x}_j)) -  \underbrace{\EX(Y_i - g_0(\bm{x}_i))}_\text{0}  =  \delta(\bm{x}_j).
\end{align}
Analogously, from equation \eqref{eq:EXOij_b}, and similarly adjusting for known $g_1(\cdot)$, the expectation equals the true individualized treatment effect for the control individual $i$
\begin{align} \label{eq:delta_ctrl}
\EX(O_{ij}^* | g_1(\bm{x}_i), g_1(\bm{x}_j)) = \underbrace{\EX(Y_j - g_1(\bm{x}_j))}_\text{0} - \EX(Y_i - g_1(\bm{x}_i))  =  \delta(\bm{x}_i).
\end{align}
Adjusting for $g_0(\cdot)$ in equation \eqref{eq:delta_trt} aims to achieve prognostic balance, which bears resemblance to prognostic score analysis \cite{hansen_prognostic_2008, nguyen_use_2019}. Conditioning on $g_1(\cdot)$ in equation \eqref{eq:delta_ctrl} is just the mirror image for $g_1(\cdot)$. In practice, $g_0(\cdot)$ and/or $g_1(\cdot)$ will of course have to be estimated and the exact equalities will become approximations. However, estimates of either one provide a matching target that is (i) one-dimensional, and (ii) is a weighted function of $\bm{x}$ aiming to retain just those elements that are required to reach $\EX(O_{ij}^*|\bm{x}_i, \bm{x}_j) = \Delta(\bm{x}_j)$ (adjusting for $g_0(\cdot)$) or $\EX(O_{ij}^*|\bm{x}_i, \bm{x}_j) = \Delta(\bm{x}_i)$ (adjusting for $g_1(\cdot)$). Hence, we implemented a 1:1 matching procedure similar to the c-for-benefit, but with two important differences. First, matching was performed based on $\hat{g}_0(\bm{x})$ as opposed to predicted treatment effect. Second, concordance was evaluated between the individual level $\hat{\delta}(\bm{x}_j)$, as opposed to the averaged $\hat{\delta}_{ij}$, and the corresponding adjusted $O_{ij}^*$'s, which fits our estimand $\theta_d$ under perfect matching on $g_0(\bm{x})$. Imperfect matching may arise from error in $\hat{g}_0(\cdot)$ and unavailable matches on the level of $\hat{g}_0(\cdot)$. We will further refer to this implementation as cben-$\hat{y}^0$. Note that a mirror image alternative could be performed when matching on $\hat{g}_1(\bm{x})$; the choice between the two might be guided by the expected quality in terms of prediction accuracy of $\hat{g}_0(\cdot)$ and $\hat{g}_1(\cdot)$, and the size of the group in which ITE predictions will be evaluated.  

\textbf{Predicted pairwise treatment effects} Recent work by van Klaveren et al.  \cite{van_klaveren_measuring_2023} and Maas et al. \cite{maas_performance_2023} suggests a modification of the c-for-benefit procedure. This novel work emphasizes the benefit of 1:1 nearest neighbour matching of treated and control cases on Mahalanobis distance, since this avoids model-dependence of the matching procedure. Furthermore, they recognize the difficulty of the original definition of predicted treatment effect for a treated-control pair (equation \eqref{eq:deltahat_ij}), and instead propose to use 'predicted pairwise treatment effects'. The latter is defined as the predicted difference in outcome risk within the matched treated-control pair (\textit{i.e.} $\hat{g}_1(\bm{x}_j) - \hat{g}_0(\bm{x}_i)$). This aligns the within-pairs observed outcome differences $O_{ij}$ and the predictions (\textit{i.e.} that now specifically target this outcome difference). However, as can be seen from equations \eqref{eq:EXOij_a} and \eqref{eq:EXOij_b}, this 'predicted pairwise treatment effects' reflects \textit{both} the treatment effect of interest \textit{and} the degree to which the model correctly predicts the within-pair difference in prognosis under the \textit{same} treatment allocation (\textit{i.e.} matching error in ${g}_0$ or ${g}_1$). Thus, it attributes correctly predicted within-pair differences in outcome risk that are unrelated to treatment to the 'predicted pairwise treatment effects'. In line, we expect this novel modification to overestimate $\theta_d$. For the remainder of this paper, we will use cben-$\hat{\delta}$ to refer to the original c-for-benefit using 1:1 matching on predicted treatment effect, and we will use cben$_{ppte}$ to refer to this recently proposed modification.

\subsection{Model-based c-statistic for benefit} \label{sec:mbcb}

Extending earlier work on model-based concordance assessment in the context of risk prediction \cite{van_klaveren_new_2016}, we propose model-based estimation of $\theta_d$ (equation \eqref{eq:theta_d}): the concordance statistic between ITE predictions and the true difference in probabilities to observe benefit between pairs of individuals. As such, model-based estimates are used to approximate $P(\Delta_k < \Delta_l)$ in equation \eqref{eq:theta_d2}. For randomly selected pairs $k,l \in 1,\ldots,n$ ($k \neq l$), and taking $Y=1$ to be harmful, there are five potential outcome configuration that signal more benefit for case $k$ than for case $l$.
\begin{enumerate}
    \item $Y_k^{a=1}=0, Y_k^{a=0}=1, Y_l^{a=1}=0, Y_l^{a=0}=0$ (benefit for $k$, no benefit for $l$)
    \item $Y_k^{a=1}=0, Y_k^{a=0}=1, Y_l^{a=1}=1, Y_l^{a=0}=1$ (benefit for $k$, no benefit for $l$)
    \item $Y_k^{a=1}=0, Y_k^{a=0}=1, Y_l^{a=1}=1, Y_l^{a=0}=0$ (benefit for $k$, harm for $l$)
    \item $Y_k^{a=1}=0, Y_k^{a=0}=0, Y_l^{a=1}=1, Y_l^{a=0}=0$ (no benefit for $k$, harm for $l$)
    \item $Y_k^{a=1}=1, Y_k^{a=0}=1, Y_l^{a=1}=1, Y_l^{a=0}=0$ (no benefit for $k$, harm for $l$).
\end{enumerate}
The corresponding probability estimates for these patterns follow easily from the model(s) for both potential outcomes. For instance, for the first pattern: $[1-\hat{P}(Y_k^{a=1}=1)] \cdot \hat{P}(Y_k^{a=0}=1) \cdot [1-\hat{P}(Y_l^{a=1}=1)] \cdot [1-\hat{P}(Y_l^{a=0}=1)]$. The sum of the five patterns is further referred to as $P_{\textnormal{benefit}, k, l}$. Subsequently, the required elements for an estimate of $\theta_d$ can be derived. From equation \eqref{eq:theta_d2}, plugging in $P_{\textnormal{benefit}, k, l}$ for $P(\Delta_k < \Delta_l)$, we estimate model-based concordance probability for benefit (mbcb) as 
\begin{equation} \label{eq:mbcb}
    \textnormal{mbcb} = \frac{\sum_k \sum_{l \neq k} \left[
    I(\hat{\delta}_k < \hat{\delta}_l) \hat{P}_{\textnormal{benefit}, k, l} + 
    \frac{1}{2} I(\hat{\delta}_k = \hat{\delta}_l) \hat{P}_{\textnormal{benefit}, k, l} \right]}
    {\sum_k \sum_{l \neq k} \left[ \hat{P}_{\textnormal{benefit}, k, l} \right]}
\end{equation}

Estimating both $\hat{\delta}(\bm{x})$ and $\hat{P}_{\textnormal{benefit}, k, l}$ from the same model ('apparent' performance), the mbcb provides $\theta_d$ for the covariate distribution used to derive the predictions and assuming that the model is correct. This is useful to evaluate the influence of case-mix on the mbcb for a given model, since c-statistics are case-mix sensitive \cite{van_klaveren_new_2016, nieboer_assessing_2016}. For actual validation of predictions $\hat{\delta}(\bm{x})$, $\hat{P}_{\textnormal{benefit}, k, l}$ should be estimated from independent data. As an example, suppose we are evaluating model $\mathcal{M}$ that can predict both $\hat{\delta}(\bm{x})$ and $\hat{P}_{\textnormal{benefit}, k, l}$ in new independent data not used to obtain $\mathcal{M}$. First, predicting both $\hat{P}_{\textnormal{benefit}, k, l}$ and $\hat{P}_{\textnormal{benefit}, k, l}$ from $\mathcal{M}$ provides $\theta_d$ for the independent data assuming that $\mathcal{M}$ is correctly specified. Second, using $\mathcal{M}$ to predict $\hat{\delta}(\bm{x})$ and an independent model based on the new data to predict $\hat{P}_{\textnormal{benefit}, k, l}$, the mbcb provides an actual estimate $\hat{\theta}_d$ of the discriminative quality of $\mathcal{M}$'s ITE predictions in the independent data.

\section{Calibration of individualized treatment effect predictions} \label{sec:calibration}

We use calibration here in the sense of the work of van Calster \cite{van_calster_calibration_2016, van_calster_calibration_2019}, Steyerberg \cite{steyerberg_clinical_2019}, and Harrell \cite{harrell_regression_2015}, where perfect calibration describes the situation where the expected outcome conditional on the prediction equals the prediction, across the range of predictions. Empirically, this involves regressing observed outcomes on model predictions and then evaluating the slope and intercept or a smooth fitted to the data. Note that these measures of calibration are sensitive to both bias and the degree of spread in the predictions.\footnote{Stevens and Poppe provide an interesting overview of different uses of the term calibration in different disciplines.\cite{stevens_validation_2020})} In case of binary outcome, calibration is typically assessed by means of logistic regression. Here we will stay with this convention, which does assume that the logistic link function is appropriate for the model under evaluation. The continuous outcome case is discussed in supplementary material \ref{app:continuous}. Regardless of the type of outcome, the challenge is to cope with the unobserved nature of the outcome of interest in case of individualized treatment effect predictions. Several methods have previously been proposed. A common descriptive method to assess individualized treatment effect calibration is to form $k$ groups based on predictions $\hat{\delta}$ and to compare the within-group observed and predicted treatment effect \cite{hoogland_tutorial_2021, van_klaveren_proposed_2018, efthimiou_measuring_2023}. While intuitive and straightforward, this approach requires arbitrary choices for split points and is often hampered by small group sizes. Also, matching based solutions have been proposed \cite{maas_performance_2023}. However, in line with the arguments for model-based calibration in outcome risk prediction \cite{van_calster_calibration_2019, crowson_assessing_2016}, we here argue for model-based calibration of individualized treatment effect predictions. Below we discuss the estimands and estimation for such a model-based approach. Note that the estimands for the alternative methods mentioned (\textit{e.g.} split-group and matching approaches) are different, and that a direct comparison of the estimators would have no clear interpretation.  

\textbf{Model-based calibration of treatment effect predictions} The main aim is to find the calibration intercept and slope for estimated treatment effects on the linear predictor scale. The anticipated intercept $\beta_0$ and slope $\beta_1$ in case of a perfect prediction are 0 and 1 respectively, as for regular prognostic model calibration \cite{steyerberg_clinical_2019, harrell_regression_2015}. Thus, the estimands of interest are regression parameters. Slopes under 1 reflect overfitting of the treatment effect predictions, and conversely, slopes over 1 reflect underfitting. Deviations of the estimated calibration intercept relate to average error in the ITE predictions (for a fixed slope). 

To cope with the unobserved outcome of interest, we first assume that control outcome risk $g_0(\bm{x})$ is known. Then, for the $n_j$ treated cases $1,\ldots,n_j$, 
\begin{align} 
\text{logit}(P(Y_j=1)) = \beta_0 + \beta_1 \hat{\delta}_{lp}(\bm{x}_j) + {g}_{lp,0}(\bm{x}_j) \label{eq:cal_model_trt}
\end{align}
where $\hat{\delta}_{lp}(\bm{x}_j) = \text{logit}(\hat{g}_1(\bm{x}_j)) - \text{logit}(\hat{g}_0(\bm{x}_j))$ and offset ${g}_{lp,0}(\bm{x}_j) = \text{logit}({g}_{0}(\bm{x}_j))$. Conversely, assuming known outcome risk in the treated $g_1(\bm{x})$, for the $n_i$ control cases $1,/ldots,n_i$,
\begin{align} 
\text{logit}(P(Y_i=1)) = -\beta_0 - \beta_1 \hat{\delta}_{lp}(\bm{x}_i) + {g}_{lp,1}(\bm{x}_i).
\label{eq:cal_model_ctrl}
\end{align}
Both can be combined for all observed outcomes $Y$ (so both arms) as
\begin{align} 
\text{logit}(P(Y=1)) = [\beta_0 + \beta_1 \hat{\delta}_{lp}(\bm{x})] (2 A_i-1) + {g}_{lp,0}(\bm{x}) A + {g}_{lp,1}(\bm{x}) (1-A).
\label{eq:cal_model_both}
\end{align}
with $A=1$ for the treated and $A=0$ for controls. In the remainder of the paper, we take the calibration intercept $\beta_0$ and calibration slope $\beta_1$ in equation \eqref{eq:cal_model_both} as our calibration estimands, conditioning on a sample $\bm{x}$ and under known $P(Y=1)$, $g_0(\bm{x})$, and $g_1(\bm{x})$. In practice, $Y$ is a noisy manifestation of $P(Y=1)$ and both $g_0(\bm{x})$ and $g_1(\bm{x})$ have to be estimated. To obtain the required estimates $\hat{\beta}_0$ and $\hat{\beta}_1$, we plug $\hat{g}_0(\bm{x})$ and $\hat{g}_1(\bm{x})$ into equation \eqref{eq:cal_model_both}.

While not the focus of the current study, the assessment of calibration of predicted individualized treatment effects is also possible without the need for estimates $\hat{\EX}(Y^{A=a} | \bm{x})$ (\textit{i.e.} $\hat{g}_0(\bm{x}$) and $\hat{g}_1(\bm{x}$). Supplementary material \ref{app:tian} introduces an approach based on the transformed covariate method by Tian et al. \cite{tian_simple_2014}, which has different estimands but a similar underlying idea.

\section{Simulation study} \label{sec:sim}
A simulation study was performed with the aim to compare performance of the different discrimination and calibration measures for ITE predictions discussed across varying sample sizes. The simulation study was performed and reported in line with recommendations by Morris et al. \cite{morris_using_2019} and using \texttt{R} statistical software version 4.2 \cite{r_core_team_r_2022}.

\subsection{Simulation study procedures}

\textbf{Data generating mechanisms:} Synthetic trial data were simulated for a trial comparing two treatments on a binary outcome. Covariates $\bm{x}_1$ and $\bm{x}_2$ were generated from independent standard normal distributions and treatment assignment was 1:1 and independent of $\bm{X}$. Data generating mechanism 1 (DGM-1) was based on a simple logistic model
\begin{align}
    \textnormal{logit}(P(Y^{A=a}_i =1)) = &-1 - 0.75 a_i + x_{i1} + 0.5 a_i x_{i2} \label{eq:dgm1}
\end{align}
DGM-1 includes main effects of treatment and $X_1$ and an interaction between treatment and $X_2$. For each of $n_{sim}=500$ simulation runs, development data sets (referred to as data sets D for development) of size 500, 750, and 1000 were randomly drawn. Validation data sets from DGM-1 were of size 1000 for each simulation run (referred to as data sets V1 for validation in data from DGM-1). Marginal event probabilities were $P(Y^{a=0}=1) \approx 0.31$ and $P(Y^{a=1}=1) \approx 0.20$. Additionally, independent validation sets of n=1000 cases were sampled from a second data generating mechanism (DGM-2) with changes in the coefficients to reflect a different population (referred to as data sets V2)
\begin{align}
    \textnormal{logit}(P(Y^{A=a}_i =1)) = &-0.5 - 0.5 a_i + 0.75 x_{i1} + 0.25 x_{i2} + 
    0.25 a_i x_{i1} + 0.25 a_i x_{i2}, \label{eq:dgm2}
\end{align}
Marginal event probabilities for the second DGM were $P(Y^{a=0} =1) \approx 0.39$ and $P(Y^{a=1} =1) \approx 0.31$. With differences in both average treatment effect and heterogeneity of treatment effect between DGM-1 and DGM-2, a model developed in a sample from DGM-1 should not perform well in individuals from DGM-2. 

\textbf{Estimands:} For discrimination, our estimand was $\theta_d$ as defined in equation \eqref{eq:theta_d2}. For calibration, our estimands were $\beta_0$ and $\beta_1$ as defined in equation \eqref{eq:cal_model_both}. 

\textbf{Methods:} The ITE model fitted to the development data was a logistic regression model estimated by means of maximum likelihood of the form
\begin{align}
    \textnormal{logit}(P(Y_i =1)) = &\beta_0 + \beta_1 a_i + \beta_2 x_{i1} + 
    \beta_3 x_{i2} + \beta_4 a_i x_{i1} + \beta_5 a_i x_{i2} \label{eq:ITEmodel}
\end{align}
Discrimination performance was assessed by means of the  original c-for-benefit (cben-$\hat{\delta}$), the c-for-benefit using 1:1 matching on predicted outcome risk under the control treatment (cben-$\hat{y}^0$), recently proposed c-for-benefit based on Mahalanobis distance matching and predicted pairwise treatment effect (cben$_{ppte}$), and the proposed model-based c-for-benefit (mbcb). Calibration performance was estimated according to equation \eqref{eq:cal_model_both}. Note that estimation of cben-$\hat{y}^0$, the mbcb, and calibration assessment require estimates $\hat{g}_0(\bm{x})$ and $\hat{g}_1(\bm{x})$. For 'apparent' evaluation, these are predictions from the ITE model. In practice, they should be based on data not used to fit the ITE model. Therefore, in bootstrap evaluations and external data simulations, $\hat{g}_0(\cdot)$ and $\hat{g}_1(\cdot)$ were estimated according to model \eqref{eq:ITEmodel} in independent samples. Performance measures were evaluated in settings (1) apparent (based on the same sample as on which the ITE model was fitted \cite{harrell_regression_2015}), (2) in interval validation using bootstrap 0.632+ adjustment (3) in interval validation using bootstrap optimism correction, (4) in external validation samples V1 generated from DGM-1, and (5) in external validation data samples V2 generated from DGM-2. A more detailed account of the procedures in available in online supplementary material \ref{app:sim_evals}.

\textbf{Performance measures:} Writing $\theta_s$ for the estimand value in simulation run $s$, and $\hat{\theta}_s$ for the corresponding estimate, performance measures were averaged across simulations $s \in 1, \ldots, n_{sim}$ in terms of root mean squared prediction error $\sqrt{\frac{1}{n_{sim}}\sum_{s=1}^{n_{sim}}(\theta_s - \hat{\theta}_s)^2}$ and visualized in terms of mean $\pm$ 1 SD for both the errors ($\theta_s - \hat{\theta}_s$) and the absolute values ($\hat{\theta}_s$). To obtain more stable estimates for calibration outcomes in presence of extreme values, those summary statistics were computed after trimming away the ten percent most extreme values.

\subsection{Discrimination results}

\begin{figure}[!htb]
\centering
\includegraphics[width=1\linewidth]{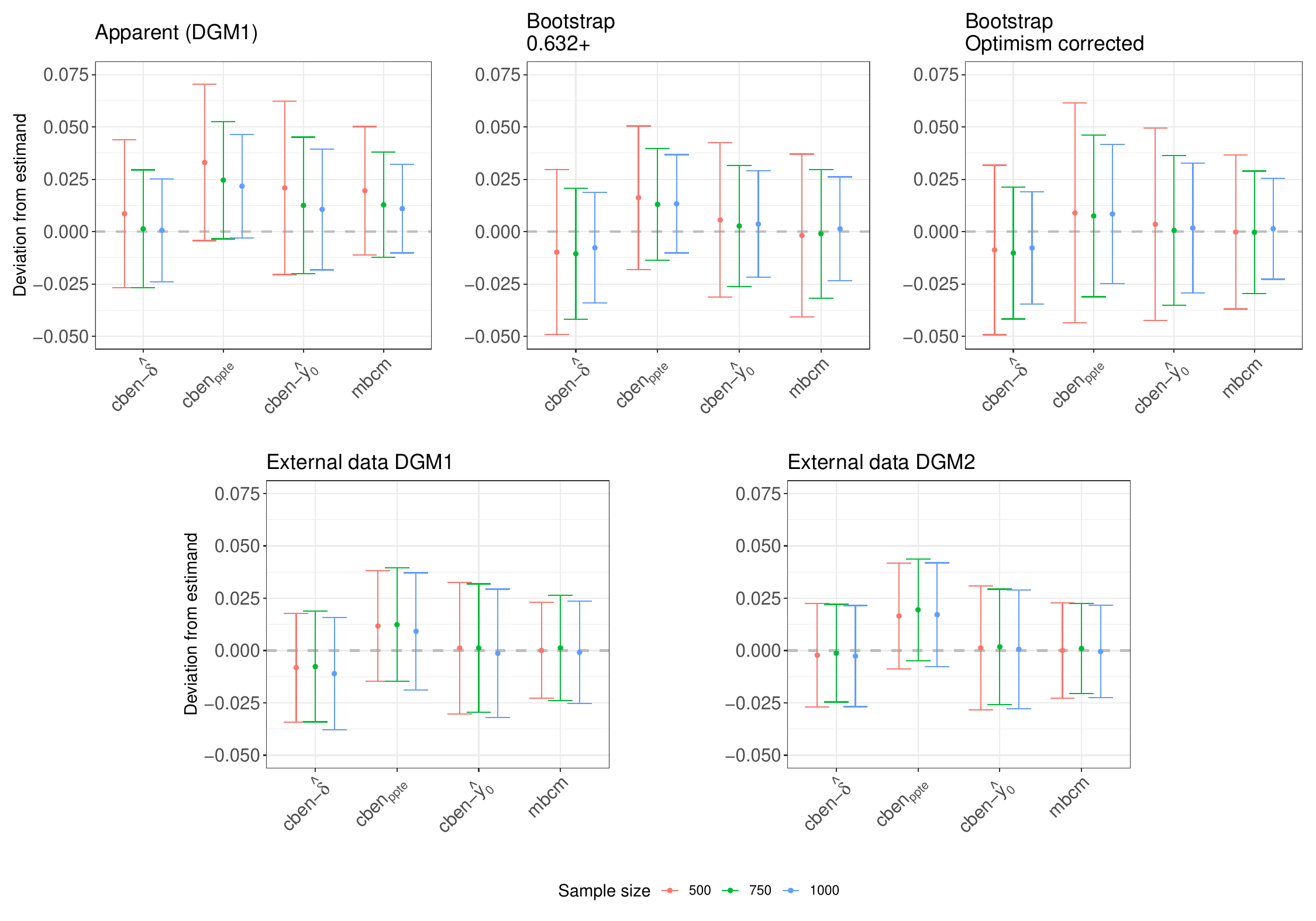}
\caption{Simulation results for the discrimination statistics in terms of mean ($\pm$ 1 SD) deviation from the estimand $\theta_d$.} 
\label{fig:simResults_discr}
\end{figure}

Figure \ref{fig:simResults_discr} (deviations from the estimands) and supplementary Figure \ref{fig:simResults_discr_suppl} (absolute value summaries) show the main simulation results with respect to the discrimination statistics. With respect to the apparent estimates, all statistics showed optimism that decreased with sample size. The original cben-$\hat{\delta}$ seemed virtually unbiased for sample sizes 750 and 1000. However, according to the results based on independent data evaluations discussed below, it was actually slightly biased downwards (in agreement with section \ref{sec:c-for-benefit_procedure}, and this canceled out the optimism here. The cben$_{optw}$ was most optimistic in agreement with section \ref{sec:cben_mods}. 

As shown in the bootstrap panels in Figure \ref{fig:simResults_discr}, both types of bootstrap evaluations successfully adjusted for optimism in the apparent evaluations. On average, bias was almost eliminated from cben-$\hat{y}^0$ and the mbcb. However, in terms of accuracy, this decrease in bias was offset by increased variability as shown by the increase in rmse between apparent and bootstrap evaluations for the best performing methods (Table \ref{tab:simResults_discr}).  

For assessment in independent validation samples from either DGM-1 or DGM-2, both cben-$\hat{y}^0$ and the mbcb were virtually unbiased, with the mbcb having the better rmse. As expected, the original cben-$\hat{\delta}$ was slightly pessimistic and the recently proposed cben$_{optw}$ was optimistic. Both had larger rmse than the mbcb.

\setlength{\tabcolsep}{2.5pt}

\begin{table}[ht]
\centering
\begin{tabular}{lrrrrrrrrrrrr} 
Statistic & \multicolumn{3}{c}{cben-$\hat{\delta}$} & \multicolumn{3}{c}{cben$_{ppte}$} & \multicolumn{3}{c}{cben-$\hat{y}^0$} & \multicolumn{3}{c}{mbcb} \\ 
  \hline   
  Sample size & 500 & 750 & 1000 & 500 & 750 & 1000 & 500 & 750 & 1000 & 500 & 750 & 1000 \\
  \hline   
 \multicolumn{2}{l}{\textbf{Development data}}&& \\
Apparent & \textbf{0.036}&\textbf{0.028}&0.025 & 0.050&0.037&0.033 & 0.046&0.035&0.031 & \textbf{0.036}&\textbf{0.028}&\textbf{0.024} \\ 
0.632+ & 0.041&0.033&0.027 & 0.038&0.030&0.027 & 0.037&0.029&0.026 & \textbf{0.039}&\textbf{0.031}&\textbf{0.025} \\ 
Opt. corrected & 0.041&0.033&0.028 & 0.053&0.039&0.034 & 0.046&0.036&0.031 & \textbf{0.037}&\textbf{0.029}&\textbf{0.024} \\ 
&&& \\
\textbf{External} &&& \\
DGM-1 & 0.027&0.027&0.029 & 0.029&0.030&0.029 & 0.031&0.031&0.031 & \textbf{0.023}&\textbf{0.025}&\textbf{0.024} \\ 
DGM-2 & 0.025&0.023&0.024 & 0.030&0.031&0.030 & 0.030&0.028&0.028 & \textbf{0.023}&\textbf{0.022}&\textbf{0.022}\\ 
   \hline
\end{tabular}
\caption{Root mean squared error against $\theta_d$ as averaged over simulation runs for each measure and for each of the sample sizes (500, 750, and 1000). Bold for best performance per setting (multiple if tied).} \label{tab:simResults_discr}
\end{table}

\subsection{Calibration results}

\begin{figure}[!htb]
\centering
\includegraphics[width=1\linewidth]{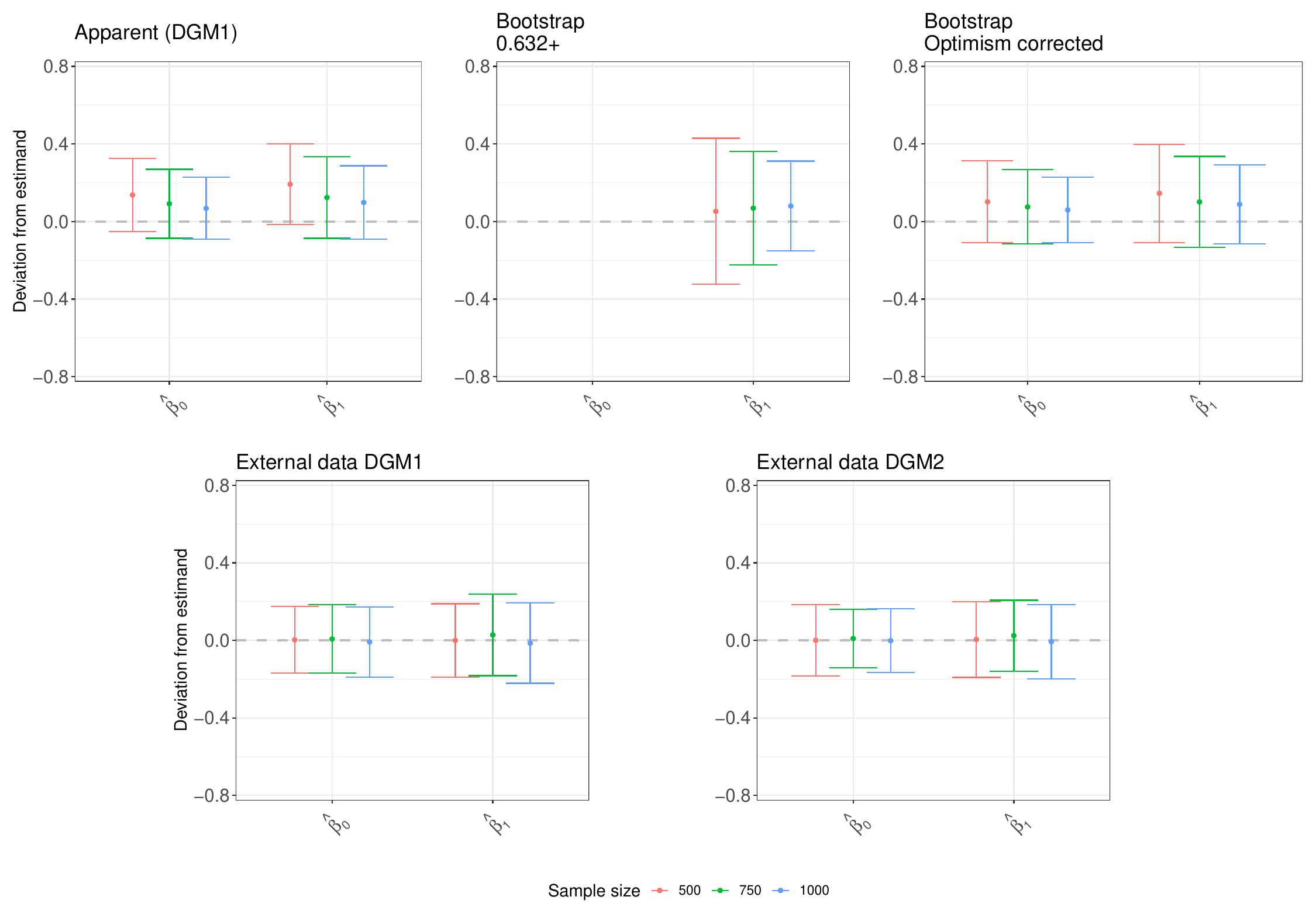}
\caption{Simulation results for the ITE calibration intercept and slope estimates. Each figures shows the deviation from the estimand (10\% trimmed mean $\pm$ 1 SD).} 
\label{fig:simResults_cal}
\end{figure}

Figure \ref{fig:simResults_cal} (deviations from the estimands) and supplementary Figure \ref{fig:simResults_cal_suppl} (absolute value summaries) show the main simulation results with respect to the calibration evaluation. In line with regular calibration of outcome risk, apparent calibration assessment is not of interest, and apparent intercepts and slopes were uniformly 0 and 1 respectively. The estimand did clearly show a decrease in overfitting with increasing sample size (Figure \ref{fig:simResults_cal_suppl}). 

Both bootstrap procedures removed some optimism from the apparent estimates and showed the decreasing risk of overfitting with increasing sample size, but were still optimistic. In fact, in terms of rmse (Table \ref{tab:simResults_cal}), the bootstrap estimates were worse than the non-informative apparent evaluation. This implies that there is not enough information in a single sample to obtain reliable out-of-sample ITE calibration estimates based on this method. This is likely related to the need to estimate $g_0(\cdot)$ and $g_1(\cdot)$ in the small number of independent out-of-sample cases. 

When validating a model in a new independent sample, calibration assessment was unbiased as desired (bottom panels Figure \ref{fig:simResults_cal}) in both DGM-1 and DGM-2, correctly identifying problems when applying the model to DGM-2. 

\begin{table}[H] 
\centering
\begin{tabular}{lrrrrrr} 
Statistic & \multicolumn{3}{c}{$\hat{\beta}_0$} & \multicolumn{3}{c}{$\hat{\beta}_1$} \\ 
  \hline   
  Sample size & 500 & 750 & 1000 & 500 & 750 & 1000 \\
  \hline   
\textbf{Development data} && \\
Apparent & 0.23& 0.20& 0.17 & 0.28& 0.24& 0.21 \\ 
0.632+ & - & - & - & 0.38& 0.30& 0.24 \\ 
Opt. corrected & 0.23& 0.21& 0.18 & 0.29& 0.26& 0.22 \\ 
&&\\
\textbf{External} && \\
DGM-1 & 0.17& 0.18& 0.18 & 0.19& 0.21& 0.21 \\ 
DGM-2 & 0.18& 0.15& 0.16 & 0.20& 0.18& 0.19 \\ 
   \hline
\end{tabular}
\caption{Root mean squared error against $\beta_0$ and $\beta_1$, after ten percent trimming, and as averaged over simulation runs for each of the sample sizes (500, 750, and 1000).} \label{tab:simResults_cal}
\end{table}

\section{Applied example: the third International Stroke Trial} \label{sec:ae}

Patients with an ischemic stroke have sudden onset of neurological symptoms due to a blood clot that narrows or blocks an artery that supplies the brain. A key component in the emergency medical treatment of these patients includes clot-busting drug alteplase (intravenous thrombolysis recombinant tissue-type plasminogen activator) \cite{powers_guidelines_2019}. 

The third International Stroke Trial (IST-3) was a randomized trial and investigated the benefits and harms of intravenous thrombolysis with alteplase in acute ischemic stroke \cite{the_ist-3_collaborative_group_benefits_2012}. This large trial included 3035 patients receiving either alteplase or placebo in a 1:1 ratio. The primary outcome was proportion of patients that was alive and independent at 6-month follow-up, which we used as outcome of interest here. Primary analyses of the treatment effect were performed with logistic regression adjusted for linear effects of age, National Institutes of Health stroke scale (NIHSS) score, time from onset of stroke symptoms to randomization, and presence (vs absence) of ischemic change on the pre-randomization brain scan according to expert assessment. This analysis showed weak evidence of an effect (OR 1.13, 95\% CI 0.95-1.35), but subgroup analyses suggested possibly heterogeneous treatment effect by age, NIHSS score, and predicted probability of a poor outcome.

For illustrative purposes, we here compare a main effects logistic regression model similar to the original adjusted analysis (model 1) with a model where all covariate-treatment interactions were included (model 2), and with random forest estimates \cite{athey_generalized_2019}. The outcome was coded as 0 for those independent and alive after 6 months and 1 otherwise. The included variables were treatment, age, NIHSS, time (from onset of stroke symptoms to randomization), and imaging status (presence vs absence of ischemic change on the pre-randomization brain scan). In the regression models, continuous variables age, NIHSS, and time, were modeled using smoothing splines. Models 2 included covariate-treatment interactions for these variables. Model1 and 2 were fitted using the \texttt{mgcv} package in \texttt{R} with default smoothing parameter selection based on generalized cross-validation \cite{wood_generalized_2017}. ITE predictions were based on the difference between potential outcome predictions under alteplase and the control condition. For the random forest predictions, control outcome risk was modeled as a regression forest in the control group with the same covariates as used in model 1 and 2 (models 3a). ITE predictions were based on a causal forest with the same covariates (models 3b). Where required for evaluation purposes, potential outcomes under the treated condition were inferred based on control predictions plus ITE predictions. Both random forests were fitted using \texttt{R} package \textbf{grf} \cite{tibshirani_grf_2023} and default settings (\textit{i.e.}, 2000 trees, honest splitting, minimum node size 5, try all variables for splits given $p<20$). All in all, this applied example illustrates different ways to assess the quality of individualized treatment effect predictions. The evaluated models were emphatically chosen for this purpose and were not developed in collaboration with clinical experts in the field. Hence, they are not meant to me applied in practice. 

The exact parameter estimates for both models are not of key interest, but the apparent performance with respect to outcome risk prediction under allocated treatment was good: c-statistics were 0.826 (model 1), 0.831 (model 2), and 0.818 (model 3a/b), with corresponding Brier scores of 0.160, 0.158, and 0.163, and Nagelkerke $R^2$ of 0.389, 0.402, and 0.370. Important to note, only the random forest implementations directly provide out-of-bag estimates for the training data, so their apparent evaluation metrics should be less optimistic. However, optimism in outcome predictions should be small given the very large $n$ to $p$ ratio. Differences between methods were small and Spearman's correlation between outcome risk predictions for the different methods were all $> 0.97$.
 
\begin{table}[ht]
\centering
\begin{tabular}{lrrrrrr}
Model & cben-$\hat{\delta}$ & cben$_{ppte}$ & cben-$\hat{y}^0$ & mbcb &  $\hat{\beta_0}$ &  $\hat{\beta_1}$  \\ 
\hline
\textbf{Apparent} &&&&& \\
M1 & 0.488 & 0.536 & 0.489 & 0.510 & -0.117 &  \\ 
M2 & 0.562 & 0.584 & 0.570 & 0.567 & 0.012 & 1.096 \\ 
M3a/b & 0.538 & 0.544 & 0.557 & 0.566 & -0.023 & 0.788 \\ 
&&&&& \\
\textbf{bootstrap 0.632+} &&&&& \\
M1 & 0.489 & 0.563 & 0.499 & 0.505 &  &  \\ 
M2 & 0.535 & 0.584 & 0.559 & 0.536 &  & 0.484 \\ 
M3a/b & 0.536 & 0.566 & 0.557 & 0.531 &  & 0.382 \\ 
&&&&& \\
\textbf{Optimism corrected} &&&&& \\
M1 & 0.485 & 0.486 & 0.475 & 0.507 & -0.118 &  \\ 
M2 & 0.534 & 0.538 & 0.544 & 0.518 & -0.047 & 0.900 \\ 
M3a/b & 0.486 & 0.459 & 0.502 & 0.537 & -0.065 & 0.486 \\
\hline
\end{tabular}
\caption{Applied example discrimination and calibration statistics for predicted individualized treatment effect.} \label{tab:aeResults}
\end{table}

The predicted ITEs, however, were very different. Model 1 predicted ITEs with median -0.020 (95\% between -0.29 and 0.00), model 2 predicted ITEs with median -0.026 (95\% between -0.161 and 0.120), and model 3b predicted ITEs with a median of -0.023 (95\% between -0.147 and 0.113). That is, the predicted treatment effect was very similar across individuals when predicted by model 1 (assuming a constant treatment effect on the log odds scale), but not when predicted by model 2 (assuming a heterogeneous treatment effect on the log odds scale) or causal random forest model 3b. Table \ref{tab:aeResults} shows the apparent and bootstrap corrected results for discrimination and calibration assessment at the ITE level for the applied example as averaged over 1000 bootstrap samples. 

With respect to ITE discrimination, both apparent and bootstrap-corrected discrimination estimates favored model 2 and 3a/b over model 1, with model 1 estimates around the no discriminative ability value of 0.5. The apparent mbcb for both model 2 and 3a/b were around 0.566 and correspond to the expected c for benefit for the covariate distribution in this sample assuming the model is correct. Bootstrap corrected results generally preferred model 2, with the exception of optimism corrected mbcb (preferring the random forest-based predictions). As a general remark, boot0.632+ corrections were quite stable, while optimism correction gave larger corrections. Based on the simulation results, their accuracy is similar in large sample. All in all, model 2 and model 3a/b are clearly better than model 1 in terms of discriminative ability. 

In terms of calibration, bootstrap corrected slope estimates suggested that both model 2 and model 3a/b were overfitted with respect to ITEs. The amount of shrinkage suggested varies considerably between the 0.632+ and optimism corrected estimates, with the 0.632+ estimate suggesting more shrinkage. Based on the simulation study, the 0.632+ was more accurate. Note that the calibration slope for model 1 is not estimable (since the ITEs have no variability on the logit scale) and the intercept estimate for model 1 clearly showed that the degree of predicted benefit was underestimated.

The results indicate that model 1 did not provide useful differentiation in terms of ITEs. While the discriminative ability of model 2 and model 3a/b seemed modest, clear benchmarks are lacking. With respect to treatment decisions, Table \ref{tab:ae_decision} shows out-of-sample treatment benefit estimates for treatment according to the model versus (i) not treated according to the model \cite{efthimiou_measuring_2023}, (ii) treatment all, and (iii) treat none are provided. All are non-significant. Nonetheless, while model 2 and model 3b are very different methods, they provided similar ITE estimates that agree with respect to sign in 80\% of cases. Also, patient characteristics of those predicted to have benefit agree with clinical knowledge \cite{powers_guidelines_2019}. For model 2 comparing the 1969 patients predicted to have benefit according to model 2 ($\hat{\delta}_{model2} < 0$) with the remaining 1066 patients ($\hat{\delta}_{model2} \geq 0$), the first were older [median(IQR) age 83 (78-87) vs 73 (63-82)], had worse symptoms [median(IQR) nihss 15(10-20) vs 6(4-9)], had less delay to randomization and thus treated earlier [median(IQR) time in hours 3.5 (2.5-4.9) vs 4.2 (3.6-4.8)], were more likely to have visual infarction on imaging (43\% vs 36\%). These figures were very similar for model 3b. Concluding, there was insufficient signal to reliably guide treatment decisions, but the proposed measures clearly differentiated between the models with and without potential (1 vs 2 \& 3a/b), and aligned with estimates of decision accuracy \cite{efthimiou_measuring_2023}. 

\begin{table}[ht]
\centering
\begin{tabular}{lrrr}
Comparison & M1 & M2 & M3b \\ 
  \hline
  Model vs opposite & -0.009 (-0.050, 0.053) & -0.031 (-0.081, 0.020) & -0.038 (-0.082, 0.010) \\ 
  Model vs treat all & 0.002 (-0.046, 0.050) & -0.010 (-0.056, 0.042)  & -0.013 (-0.061, 0.038)  \\ 
  Model vs treat none & -0.013 (-0.055, 0.044) & -0.025 (-0.071, 0.027) & -0.027 (-0.073, 0.020) \\ 
   \hline
\end{tabular}
\caption{Applied example: out-of-sample benefit estimates of treating according to the model (treat is benefit < 0, do not treat otherwise. Median and 95\% bootstrap interval are shown.} \label{tab:ae_decision}
\end{table}

\section{Software} \label{sec:software}
\texttt{R} package \textbf{iteval} (\url{https://github.com/jeroenhoogland/iteval}) provides an implementation of the cben-$\hat{\delta}$, cben-$\hat{y}^0$, mbcb, and calibration measures as defined in this paper in the freely available \texttt{R} software environment for statistical computing \cite{r_core_team_r_2022}. The cben$_{ppte}$ has been implemented in \texttt{R} package \textbf{HTEPredictionMetrics} available on GitHub (\url{https://github.com/CHMMaas/HTEPredictionMetrics})\cite{maas_performance_2023}. 

\section{Discussion} \label{sec:discussion}

Measures of calibration and discrimination have a long history in the context of prediction models for observed outcome data, especially of the binary type. However, the evaluation of prediction models for individualized treatment effects (ITE) is more challenging due to the causal nature of the predictions and the resulting unobservable nature of individualized treatment effects. In this paper, we used the potential outcomes framework \cite{rubin_causal_2005} to gain insight into existing performance measures \cite{van_klaveren_proposed_2018, maas_performance_2023, van_klaveren_measuring_2023}, clearly defined the estimands of interest, and developed model-based measures of discrimination and calibration for ITE prediction models. The model-based proposals avoid the need for matching and at the same time avoid the bias associated with existing measures, and are applicable for all prediction methods that can provide predictions for both potential outcomes. Measures of discrimination provide insight into the degree to which the model correctly ranked the predicted treatment benefit, while measures of calibration provide information on bias and over/underfitting. While the primary focus was on dichotomous outcomes, we also provided residual-based approaches for continuous outcome models. As such, our work provides generally applicable tools for the endeavor of evaluating ITE prediction models in randomized data.

In terms of discriminative ability, the proposed model-based c-for-benefit (mbcb) provides both a normal performance measure and an expected (case-mix adjusted) reference level for new data (in line with the model-based c-statistic \cite{van_klaveren_new_2016}. The latter is relevant because concordance probabilities are known to be sensitive to case-mix \cite{nieboer_assessing_2016}. Also, bootstrap procedures were proposed that adjust for optimism when no external data are available. In the simulation study, the mbcb estimates were best in terms of both bias and root mean square error across settings. The original cben-$\hat{\delta}$ \cite{van_klaveren_proposed_2018} has a more difficult interpretation and was downward biased in the simulation study, but was very stable throughout. In contrast, the recent cben$_{ppte}$ \cite{maas_performance_2023, van_klaveren_measuring_2023} was consistently optimistic. Our adaptation of the matching-based c-for-benefits (the cben-$\hat{y}^0$) removed the bias, but at too high a cost in terms of variability. We hypothesize that the stability of the mbcb is due to the lack of need for a matching algorithm. 

In terms of calibration, the potential outcomes framework provided a model-based method for evaluating ITE prediction calibration. Again, the model-based nature avoided the need for matching algorithms, and the results of the simulation study showed unbiased estimation in independent validation data. However, the main proposal depends on the accuracy of the underlying potential outcome predictions under both treatment conditions. Thus, misspecification of the potential outcome models may invalidate the proposed calibration measure. While this may seem like a significant cost, we emphasize that the medical decision making context for which these models were developed requires accurate potential outcome predictions in conjunction with ITE predictions. Therefore, we consider accurate outcome prediction models to be a prerequisite for ITE modeling, and their evaluation should be a principle part of performance evaluation in practice \cite{steyerberg_clinical_2019, harrell_regression_2015}. Nonetheless, for use cases where only the ITEs are relevant, we proposed an alternative calibration method based on the work Tian et al. \cite{tian_simple_2014} that does not depend on a potential outcome model.

A key finding for both the ITE discrimination and calibration measures was that bootstrapping procedures were able to remove optimism (i.e., reduce bias), but that the increase in variance of the estimator generally led to increased root mean squared error compared to apparent evaluation. This implies that external data are needed to accurately evaluate ITE predictions. The underlying reason is the need for accurate potential outcome predictions based only on the out-of-sample cases, for which the 36.8\% of out-of-sample cases in a bootstrap procedure were apparently insufficient. Nevertheless, bootstrap procedures are still to be preferable to apparent assessment when nothing else is available, as they provide a fairer estimate on average. Also, there was little overfitting in our examples due to large $n$ to $p$ ratios, but optimism in apparent estimates can be much more severe and can vary widely between methods. 

\textbf{Related work} The key literature underlying the developments in this paper mainly arose from the fields of (medical) statistics and epidemiology, but there are recent and connected developments in econometrics and machine learning, some of which venture beyond randomized data and the large $n$ to $p$ ratio in this paper. While ITE estimation was not the focus of our work, it is worth discussing how it relates to recent machine learning methods. In particular, recent years have seen a rapid growth in regression and machine learning methods that aim to estimate individualized treatment effects directly, without estimating the potential outcomes themselves. These methods benefit in settings where the functional form of the treatment effect is less complex than the response surfaces of the potential outcomes, and thus easier to learn or model. Popular methods include transformed covariate regression \cite{tian_simple_2014} and in particular tree-based methods \cite{hill_bayesian_2011, athey_recursive_2016, wager_estimation_2018, athey_generalized_2019, hahn_bayesian_2020}. In addition, there has been much interest in meta-learners that decompose the estimation of individualized treatment effects into separate prediction problems that can be approached using any prediction method (e.g., regression, machine learning) \cite{chernozhukov_generic_2018, kunzel_metalearners_2019, 
nie_quasi-oracle_2021, imai_statistical_2022, kennedy_towards_2023}. Typically, these methods require prior estimation of the potential outcomes and propensity scores, which are then combined into transformed outcomes that can be regressed on the covariates to predict individual treatment effects. Prominent examples include the X learner \cite{kunzel_metalearners_2019}, the DR learner \cite{kennedy_towards_2023}, and the R learner \cite{nie_quasi-oracle_2021}. Extensive simulation studies investigating the properties of these causal machine learning methods are available elsewhere \cite{knaus_machine_2021, jacob_cross-fitting_2020, okasa_meta-learners_2022}.

What all of these methods have in common is that they either do not estimate potential outcomes, or treat them as nuisance parameters that are used to obtain more accurate personalized treatment effect estimates. Also, they either inherently provide shrinkage and selection techniques, or can build on methods that do, and thus have an advantage in settings with a higher tension between model complexity and sample size (e.g., high-dimensional settings). This has been shown to be beneficial in terms of the accuracy of predicted ITEs, particularly for observational data and strong treatment selection \cite{kennedy_towards_2023, okasa_meta-learners_2022}. However, if the potential outcome predictions are of interest, separately predicting the potential outcomes and individualized treatment effects leads to the unwanted situation that $\hat{\EX}(Y^{A=1} | \bm{x}_i) - \hat{\EX}(Y^{A=0} | \bm{x}_i) \neq \hat{\delta}(\bm{x}_i)$. The pragmatic solution taken in our applied example was to separately estimate $\hat{\EX}(Y^{A=0} | \bm{x})$ and $\hat{\delta}(\bm{x})$ and infer $\hat{\EX}(Y^{A=1} | \bm{x})$ from them. While this leads to consistency between predicted potential outcomes and ITEs, it does not use the most accurate (directly estimated) model for $\hat{\EX}(Y^{A=1} | \bm{x})$. 

With respect to the proposed measures of discrimination and calibration for individualized treatment effects, the discrimination estimand formulated in section \ref{sec:discrimination} is not within reach for methods estimating individualized treatment effects $\delta(\bm{x})$ without informing on $\hat{\EX}(Y^{A=a} | \bm{x})$. This is because the probability of observing a benefit for an individual, or observing a difference in benefit between two individuals, depends on the risk in the absence of treatment. In general, medical decision making for individuals is difficult when only the treatment effect estimate is available, and not the outcome prediction under a reference condition. Obvious applications that depend only on $\delta(\bm{x})$ are beyond the scope of our work, but include ranking groups according to their predicted benefit for judicious allocation of limited resources. Research specifically aimed at evaluating a model's prioritization qualities includes recent work on prioritization rules via rank-weighted average treatment effects \cite{yadlowsky_evaluating_2021}. With respect to calibration, we have proposed an alternative method that does not depend on the potential outcome model(s), but it requires further study and comparison with a recent arXiv paper by Xu and Yadlowsky \cite{xu_calibration_2022} that provides a fairly general calibration metric for methods that only provide ITEs. Their proposal examines the $\ell_p$ norm of the expected calibration error for predicted treatment effect heterogeneity using nonparametric methods, which have different estimands compared to our proposal \cite{xu_calibration_2022}. Also, Chernozhukov has proposed an estimation framework that targets key features of individualized treatment effect, with a particular interest in approximation based on a linear function of proxy ITE predictions from auxiliary data \cite{chernozhukov_generic_2018}. Although intended for estimation and inference purposes and, unlike the current work, based on transformed outcome models, their estimation objective bears resemblance to our calibration objective without assuming a specific parametric form.

\textbf{Limitations} Limitations of the current work include the relatively narrow scope of the simulation study, which was conducted primarily for illustrative purposes.We also limited our use case to settings with randomized data and models that provide individualized treatment effect estimates as the difference between two potential outcome predictions. In addition, we focused on parametric measures of discrimination and calibration in line with the existing literature and the common context of very limited sample size, but recently proposed nonparametric measures are promising \cite{xu_calibration_2022}. Important questions remain regarding the relationship between discrimination and calibration at the outcome and ITE levels, and the relationship between discrimination and calibration statistics and the clinical utility of the models. With respect to uncertainty estimates, bootstrap procedures provide a viable option.

\textbf{Future work} In terms of future research, it would be interesting to evaluate whether some level of grouping is beneficial for evaluating model performance. Paradoxically, the goal of precision underlying the development of ITE models may hinder the ability to evaluate them, since individual-level treatment effects are inherently unobservable. In large sample situations, it would also be interesting to allow for a more flexible estimation of the calibration slope beyond the current linear implementation, which would allow for the construction of E-statistics and an integrated calibration index \cite{maas_performance_2023, austin_integrated_2019}.

\textbf{Conclusion} In summary, we have provided a principled review of existing measures of discrimination and calibration for models predicting individualized treatment effects, and proposed model-based methods that avoid the need for matching and reduce bias. Further research is needed to improve understanding of the precise properties of these measures under different conditions of sample size, degree of treatment effect heterogeneity, and explained variation, and to explore the relationship with novel estimators for related estimands in the machine learning literature.  

\section*{Acknowledgements}
This project received funding from the European Union’s Horizon 2020 research and innovation program under ReCoDID grant agreement No 825746. Jeroen Hoogland and Thomas P. A. Debray acknowledge financial support from the Netherlands Organisation for Health Research and Development (grant 91215058). Thomas P. A. Debray also acknowledges financial support from the Netherlands Organisation for Health Research and Development (grant 91617050). Orestis Efthimiou was supported by the Swiss National Science Foundation (Ambizione grant number 180083). We like to thank the researchers involved in the original stroke trial for use of their data \cite{the_ist-3_collaborative_group_benefits_2012, sandercock_p_third_2016}.

\section*{Data Availability Statement}
Data for the International Stroke Trial-3 applied example are publicly available \cite{sandercock_p_third_2016}. \texttt{R} package \textbf{iteval} is available on GitHub (\url{https://github.com/jeroenhoogland/iteval}) and provides functions to derive the cben-$\hat{\delta}$, cben-$\hat{y}^0$, mbcb, and calibration measures as defined in this paper. Github repository \textbf{iteval-sims} (\url{https://github.com/jeroenhoogland/iteval-sims}) provides the required files and instructions for replication of the simulation study.

\bibliographystyle{ieeetr}

\renewcommand\thefigure{\thesection.\arabic{figure}}    
\setcounter{figure}{0} 

\renewcommand\thetable{\thesection.\arabic{table}}    
\setcounter{table}{0} 


\appendix
\numberwithin{equation}{section} 

\section*{Supplementary Material}

\section{Binomial outcome data} \label{app:binom_challenges}

\subsection{Absolute risk, risk difference, and binomial error}
Focusing on binary outcomes, assume we observe outcome $Y_i \in \{0,1\}$ and covariate status $\bm{x}_i$ for each individual $i$. Using data on $n$ individuals, we can model the outcome risk $P(Y_i = 1 | A=a_i, \bm{X}=\bm{x}_i)$. There are two sources of error when using such a model to predict binary outcomes. There is the reducible error in modeling the risk (\textit{i.e.}, how well the modelled probability approximates the actual probability of an event), and there is the irreducible error in the difference between the actual probability of an event and its manifestation as a $\{0,1\}$ outcome (binomial error). 

Adding to this, actual interest is in the difference in outcome risk under different treatment assignment $a \in \{0,1\}$. That is, interest is in $p(Y_i = 1|A = 1, \bm{X}=\bm{x}_i) - p(Y_i = 1|A = 0, \bm{X}=\bm{x}_i)$. The range of possible true (and estimated) treatment effects (risk differences) includes all values in the $[-1,1]$ interval, but the observed difference between any two outcomes can only be one of $\{-1, 0, 1\}$. An example may be helpful to appreciate the large influence of irreducible error in this setting. For instance, regardless of any modeling, assume that an active treatment (as compared to a control condition) reduces outcome risk from 25\% to 20\% for a certain individual. Moreover, assume that these probabilities are known exactly and that this individual can be observed under both treatment conditions. A simple probabilistic exercise\footnote{For instance, $P(Y^0=0, Y^1=0)=(1-P(Y^0=0))(1-P(Y^1=0))=(1-0.25)(1-0.2)=0.6$} shows that the different outcome probabilities are $P(Y^0=0, Y^1=0)=0.6, P(Y^0=0, Y^1=1)=0.15, P(Y^0=1, Y^1=0)=0.2$, and $P(Y^0=1, Y^1=1)=0.05$. That is, the probability that the active treatment induces any observed outcome difference is 35\%, and only 20\% is in the expected direction (\textit{i.e.} in the direction of the treatment effect). This is just due to the \textit{irreducible} error, apart from any modeling issues, and ignoring the fact that in practice only one potential outcome is observed of each individual. The insensitivity of binary endpoints is of course well known in the context of trials, where a larger number of replications can provide a solution when the average treatment effect is of interest. In the case of individualized treatment effect estimation however, the required number of replications is more challenging to control due to its complex dependence on all individual-level characteristics of interest. 

\subsection{Scale matters}
Models that predict the risk of a binary event commonly make use of a link function in order to map a function of the covariates in $\R$ onto the probability scale \cite{agresti_categorical_2013}. Such link functions, such as the logit or inverse Gaussian, are inherently non-linear and hence do not preserve additivity. Consequently, a treatment effect that is constant (\textit{i.e.}, does not vary with other covariates) before applying the link function \textit{shall} vary with other covariates on the risk scale and vice versa. As an example, we write $h^{-1}$ for an inverse link function and take control risk to be a function $f(\cdot)$ of only one random variable $X$ (\textit{i.e.}, $P(Y^{a=0} =1|X=x) = h^{-1}(f(X))$). Subsequently, assume a constant (homogeneous) relative treatment effect $d$ such that $P(Y^{a=1} =1|X=x)=h^{-1}(f(X) + d)$, then the absolute treatment effect necessarily depends on $X$, since
\begin{align} \label{eq:abs-rel}
  \delta(x) &= P(Y^{a=1} =1|X=x) - P(Y^{a=0} =1|X=x) \\ \nonumber
    &= h^{-1}[f(X) + d] - h^{-1}[f(X)] \neq h^{-1}[f(X) + d - f(X)] = h^{-1}(d)
\end{align}
unless $h^{-1}(\cdot)$ is linear. Consequently, between-individual variability (\textit{i.e.}, variability in terms of $X$) directly changes control outcome risk \textit{and} affects the absolute effect of $d$ on the probability scale even if $d$ is constant. For instance, a constant treatment effect on the log-odds scale translates into heterogeneous treatment effect on the risk difference scale. Thereby, relatively simple treatment effect structures may lead to meaningful between-individual treatment effect variability at the risk difference level if there is large variability in $h^{-1}[f(X)]$ \cite{hoogland_tutorial_2021, harrell_viewpoints_2018}. In addition, treatment effect may interact with $X$ in the domain of $h^{-1}(\cdot)$, \textit{i.e.}, we may directly model treatment effect heterogeneity. These two sources of variability in $\delta(x)$ can no longer be discerned when evaluating just the estimates $\hat{\delta}(x)$. Hence, the benefit in terms of interpretation of measures on the scale of $\delta(x)$ \cite{murray_patients_2018}, as of interest in this paper, has a price in that they conflate variability in $\hat{\delta}(x)$ from different sources: between-subject variability in $P(Y^{a=0} =1|X=x)$ and genuine treatment effect heterogeneity on the scale used for modeling. 

\section{Measures of association for ordinal variables} \label{app:history}

The c-statistic used in the main text to measure discriminative performance has its origins in earlier work dating back to Kendall's proposal of $\tau_a$, $\tau_b$ and $\tau_c$ \cite{kendall_new_1938, kendall_rank_1970, stuart_estimation_1953}, Goodman and Kruskal's $\gamma$ \cite{goodman_measures_1954}, and Somers's $d_{xy}$ and $d_{yx}$ \cite{somers_new_1962}. All of these measures attempt to quantify a monotone relationship between two variables that have a natural order, but they have different properties. This section of the Supplementary Material provides a brief overview of these association measures that led to the formulation of the c-statistic in the main text. 

What all measures have in common is that they are fractions with a common numerator that have been written if different but equivalent forms. For variables $x$ and $y$ observed on $1,\ldots,n$ individuals, for all fully ranked pairs, the numerator can be written as

\begin{align*}
    & \text{Pr(concordance)} - \text{Pr(discordance)} \\    
\end{align*}   

with, for pair $i,j$, concordance defined as either $x_{i} < x_{j} \; \& \; y_{i} < y_{j}$ or $x_{i} > x_{j} \; \& \; y_{i} > y_{j}$ and discordance defined as either $x_{i} < x_{j} \; \& \; y_{i} > y_{j}$ or $x_{i} > x_{j} \; \& \; y_{i} < y_{j}$. Ties arise when $x_{i} = x_{j}$ or $y_{i} = y_{j}$. Equivalently, the numerator can be written as 

\begin{align*}
    & \mathop{\sum \sum}_{i<j} {\text{sign}(X_i - X_j)\text{sign}(Y_i - Y_j)}
\end{align*}   

for all pairs $i,j$. Difference between the association measures arise in the denominator and relate to the handling of ties and the symmetric or asymmetric handling of $x$ and $y$.

\subsection{Symmetric measures of association}

In the original proposal for $\tau$ \cite{kendall_new_1938} (now known as $\tau_a$), the denominator is the total number of pairs $\binom{n}{2}$.\footnote{This allows for an alternative notation as $\tau_a = \EX \{ {\text{sign}(X_i - X_j)\text{sign}(Y_i - Y_j)} \}$.} Hence, $\tau_a$ expresses the proportionate excess of concordant over discordant pairs among all pairs. In absence of ties, it ranges from -1 (perfect discordance) to +1 (perfect concordance), but these bounds are closer to zero in case of ties. To adjust for ties, the denominator of $\tau_b$ equals the geometric mean of the number of non-tied pairs on X and the number of non-tied pairs on Y \cite{kendall_rank_1970}:

\begin{align*}
    & \tau_b =  \frac{\mathop{\sum \sum}_{i<j} {\text{sign}(X_i - X_j)\text{sign}(Y_i - Y_j)}}{
                      \sqrt{\mathop{\sum \sum}_{i<j} {\text{sign}(X_i - X_j)^2}
                      \mathop{\sum \sum}_{i<j} {\text{sign}(Y_i - Y_j)^2}}}
\end{align*}   

This $\tau_b$ is widely used in software implementations (\textit{e.g.}, using \texttt{cor()} in \texttt{R} with method \texttt{method = "Kendall"}). The later $\tau_c$ was motivated by the fact that $\tau_b$ can still not achieve $\pm 1$ in case of ties, as is directly clear from the Cauchy inequality \cite{stuart_estimation_1953}. The denominator for $\tau_c$ is $\frac{1}{2} n^2 (m-1)/m$, with $m$ the longest diagonal in a $x,y$ contingency table, which gets closer to the $\pm 1$ bounds than $\tau_b$ in general, and exactly to $\pm 1$ when $n$ is a multiple of $m$. Nonetheless, neither of $\tau_b$ and $\tau_c$ have an easy interpretation in words. 

As noted by Somers \cite{somers_new_1962}, Goodman and Kruskal's $\gamma$ \cite{goodman_measures_1954} does have a straightforward interpretation. It yet again has the name numerator, but the denominator is simply $\text{Pr(concordance)} + \text{Pr(discordance)}$. It is easily verified that it may reach $\pm 1$ in case of ties and gives the proportionate excess of concordant over discordant pairs among all pairs which are fully discriminated, or fully ranked \cite{somers_new_1962}. 

\subsection{Asymmetric measures of association}
The above measures treat $x$ and $y$ symmetrically; that is, they make no distinction between dependent and independent variables. Somers proposed the asymmetric variants $d_{xy}$ and $d_{yx}$ \cite{somers_new_1962}. Again, they have the same numerator. The denominator of $d_{yx}$ is the number of pairs not tied on X, and conversely the denominator of $d_{xy}$ is the number of pairs not tied on Y. So for $d_{xy}$,

\begin{align*}
    & d_{xy} =  \frac{\mathop{\sum \sum}_{i<j}  {\text{sign}(X_i - X_j)\text{sign}(Y_i - Y_j)}}{
                      \mathop{\sum \sum}_{i<j} {\text{sign}(Y_i - Y_j)^2}} \\         
\end{align*}   

It reflects the proportionate excess of concordant over discordant pairs among all pairs which are not tied on $Y$. When sampling pairs at random from a bivariate distribution, it can also be interpreted as the difference in probability between concordant and discordant pairs, conditioning on the fact that ties on the independent variable are ignored \cite{somers_new_1962}. 

Now the c-statistic as first described by Harrell \cite{harrell_evaluating_1982} is a simple transformation of Somers $d_{xy}$, where $c = d_{xy} / 2 + \frac{1}{2}$, hence ranging from 0 to 1. The novelty of the proposed c-statistic was to apply this measure to right-censored survival data and to extend the notion of ties on the outcome $Y$ to pairs that are incomparable due to censoring (\textit{i.e.}, pairs where both cases are censored, or pairs where one case had an event at a time later than the censoring time of the censored case in the pair). Instead of a transformation of $d_{xy}$, it can also be written as

\begin{align*}
    & c =  \frac{\mathop{\sum \sum}_{i<j} \{ I[\text{sign}(X_i - X_j)\text{sign}(Y_i - Y_j)=1] + 
                \frac{1}{2} I[\text{sign}(X_i - X_j)=0 \; \& \; \text{sign}(Y_i - Y_j) \neq 0] \}}{
                      \mathop{\sum \sum}_{i<j} {\text{sign}(Y_i - Y_j)^2}} \\         
\end{align*}   

where $I[\text{sign}(X_i - X_j)\text{sign}(Y_i - Y_j)=1]$ counts concordant pairs, $\text{sign}(Y_i - Y_j)^2$ counts pairs not tied on $Y$, and $\frac{1}{2} I[\text{sign}(X_i - X_j)=0 \; \& \; \text{sign}(Y_i - Y_j) \neq 0]$ ensures that pairs tied for $X$ bring the c-statistic closer to the neutral value 0.5. Note that for Somers' $d_{xy}$, specific handling of pairs tied on $X$ is not required since its neutral value is 0. In terms of interpretation, the c-statistic reflects the proportion of concordant pairs amongst pairs not tied on the outcome. A c-statistic of 1 indicates complete concordance, 0 indicates complete discordance, and 0.5 indicates no association. A probabilistic interpretation for a randomly sampled pair is the probability of concordance given that the pair is untied on the outcome.

\section{Continuous outcome data} \label{app:continuous}

While measures of discrimination and calibration have found their origin in research on dichotomous outcomes, they may also be of value for the assessment of ITE models for continuous outcomes. This supplementary section describes model-based variants for continuous outcomes. 

\subsection{Discrimination}

The model-based c-statistic as introduced for binary outcome data specifically evaluates the pairwise concordance probability between predicted treatment effect as a risk difference and the probability to observe benefit on the outcome. This choice was made to stay as close as possible to the c-for-benefit as proposed by van Klaveren et al. \cite{van_klaveren_proposed_2018}, which evaluates the pairwise concordance probability between predicted treatment effect and the difference in observed outcomes between matched cases and controls. For treatment effects measured on continuous outcomes, such a probabilistic approach is not required. The estimand can be written in form analogous to equation \eqref{eq:theta_d2} as
\begin{align} \label{eq:theta_d_cont}
\theta_d^{cont} &= \frac{
    P(\delta_k < \delta_l \cap \hat{\delta}_k < \hat{\delta}_l) + 
    \phi^*}{
    P(\delta_k < \delta_l)},
\end{align}
where $\phi^* = \frac{1}{2} P(\delta_k < \delta_l \cap \hat{\delta}_k = \hat{\delta}_l)$. As such, it shared the interpretation that for a randomly selected control-treated pair, $\theta_d^{ct}$ reflects the probability that the ITE predictions will be concordant with the true treatment effects. A possible estimator for the case where independent data are available to estimate the $\theta_d^{ct}$, we define 
\begin{equation} \label{eq:mbcb_cont}
    \textnormal{mbcb}_{cont} = \frac{\sum_k \sum_{l \neq k} \left[
    I(\hat{\delta}_{o,k} < \hat{\delta}_{o,l}) I(\hat{\delta}_{n,k} < \hat{\delta}_{n,l}) + 
    \frac{1}{2} I(\hat{\delta}_{o,k} = \hat{\delta}_{o,l}) I(\hat{\delta}_{n,k} = \hat{\delta}_{n,l}) \right]}
    {\sum_k \sum_{l \neq k} \left[ I(\hat{\delta}_{n,k} = \hat{\delta}_{n,l}) \right]},
\end{equation}

where subscripts $o$ and $n$ denote predictions based on the original model and the newly fitted model in the independent data respectively. Note that evaluation of the $\textnormal{mbcb}_{cont}$ for a single model (\textit{i.e.} apparent performance) is not meaningful and trivially equal to 1. When validating a model in independent data, one would expect to find a $\textnormal{mbcb}_{cont} \leq 1$, with values close to 1 expressing high concordance and values close to 0.5 expressing a total lack of concordance. In absence of ties, $\textnormal{mbcb}_{cont}$ is just Kendall's $\tau$ re-scaled to the [0,1] range (\textit{i.e.}, $2(\textnormal{Kendall's } \tau - \frac{1}{2})$). Incorporation of the predicted probabilities to actually observe benefit in random pairs $j,k$ (denoted by $P_{\textnormal{benefit}, k, l}$ in the main text) would result in a less trivial metric, but is beyond our current scope. 

\subsection{Calibration}

Calibration for ITE models with a continuous outcome and an identity link function can be seen as a simplification of the case for dichotomous outcomes. In such a linear model, treatment effects on the outcome level only depend on terms that include the potential treatment assignment. 

Analogous to the dichotomous case, we approximate observed treatment effect by adjustment of the observed outcomes for the prediction under the control condition. For continuous outcomes in the treated $Y_j$, these are residuals $Y_j - \hat{g}_0(\bm{x}_j)$. Subsequently, these residuals can be regressed on $\hat{\delta}(\bm{x}_j)$. That is,
\begin{align} 
Y_j - \hat{g}_0(\bm{x}_j) = \beta_0 + \beta_1 \hat{\delta}(\bm{x}_j) + \epsilon_j
\end{align}
for individuals $j \in 1,\ldots,n_j$ and with $\epsilon_j \sim N(0,\sigma^2)$. In the same line, for controls $i$, we have
\begin{align} 
\hat{g}_1(\bm{x}_i) - Y_i = \beta_0 + \beta_1 \hat{\delta}(\bm{x}_i) + \epsilon_i
\end{align}
for individuals $i \in 1,\ldots,n_i$ and with $\epsilon_i \sim N(0,\sigma^2)$.
The anticipated intercept, slope and their interpretation is the same as in the main text. Since the right hand side in both equations is identical, they can be estimated at once to increase accuracy by (\textit{e.g.} by $\sqrt{2}$ for 1:1 allocation). In addition to model-based evaluation, a smooth curve such as a loess (locally estimated scatterplot smoothing) estimate can be drawn through a scatterplot of $Y_j - \hat{g}_0(\bm{x}_j)$ and/or $\hat{g}_1(\bm{x}_i) - Y_i$ versus $\delta(\bm{x}_j)$ to provide a visual evaluation of ITE calibration for continuous outcomes.

\section{Performance evaluation details} \label{app:sim_evals}

\subsection{Apparent performance}
For apparent performance was evaluation, the ITE model was evaluated on the same samples in which it was fitted, directly evaluating its predictions $\hat{\delta}(\bm{x})$ for these samples and plugging in the ITE models estimates of $\hat{g}_0(\bm{x})$ and $\hat{g}_1(\bm{x})$. Apparent performance estimates were implemented as a check of procedures and to show optimism, and are in general not recommended. In fact, for calibration assessment of ITE models based on maximum likelihood estimation, note that the estimates will invariably be $\hat{\beta}_0=0$ and $\hat{\beta}_1=1$, since the calibration model is of the exact same type. 

\subsection{Internal validation}
\textit{Discrimination} \newline
Internal validation was performed based on a nonparametric bootstrapping procedure based on 100 bootstrap samples. Performance estimates were based on either a 0.632+ method  \cite{efron_improvements_1997} adapted for application in the context of c-statistics or on optimism correction \cite{harrell_regression_2015}. 

The adapted 0.632+ method provides a weighted average of apparent performance and average out-of-sample performance as based on predictions from bootstrap models for the cases not in the bootstrap sample. Writing $\hat{c}_{app}$ (scalar) for the apparent c-statistic and $\hat{c}_{oos}$ (scalar) for the average out-of-sample c-statistic across bootstrap replications, 

\begin{align}
  \hat{c}_{oos} &= \left\{ \begin{array}{l}
              \textnormal{min}(\gamma, \hat{c}_{oos}), \quad \hat{c}_{app} \geq \gamma \\
              \textnormal{max}(\gamma, \hat{c}_{oos}), \quad \hat{c}_{app} < \gamma \\
            \end{array} \right. \\ 
  R &= \left\{ \begin{array}{l}
              \frac{|\hat{c}_{app} - \hat{c}_{oos}|}{|\hat{c}_{app} - \gamma|}, \quad |\hat{c}_{oos} - \gamma| < |\hat{c}_{app} - \gamma| \\
              0, \quad \textnormal{otherwise} \\
            \end{array} \right. \\ 
  w &= \frac{0.632}{1 - 0.368 R} \\
  \hat{c}_{0.632+} &= \hat{c}_{app}(1-w) + w \hat{c}_{oos}
\end{align}

where $\gamma$ is the value of the statistic for an uninformative model (so $\gamma=0.5$ for c-statistics), and $w$ is a weight that depends on the discrepancy between apparent and out-of-sample performance. To prevent that $R$ falls outside of the $(0,1)$, we avoid the possibility of bootstrap correction towards a point beyond the no information threshold by replacement of $\hat{c}_{oos}$ with $\hat{c}_{oos}'$ throughout, with 

\begin{align}
  \hat{c}_{oos}' &= \left\{ \begin{array}{l}
              \textnormal{min}(\gamma, \hat{c}_{oos}), \quad \hat{c}_{app} \geq \gamma \\ \label{eq:coos_prime}
              \textnormal{max}(\gamma, \hat{c}_{oos}), \quad \hat{c}_{app} < \gamma \\
            \end{array} \right. \\ \nonumber
\end{align}

Subsequently, $R$ reflects the degree of overfitting and ranges from zero to one, with $w$ depending only on $R$ and ranging from 0.632 and 1. Thereby, $\hat{c}_{0.632+}$ moves towards $\hat{c}_{oos}'$ when the amount of overfitting ($|\hat{c}_{app} - \hat{c}_{oos}'|$) is large with respect to the models gain relative to no information ($|\hat{c}_{app} - \gamma|$). The choice to use $\hat{c}_{oos}'$ instead of $\hat{c}_{oos}'$ in \eqref{eq:coos_prime} was to avoid correction of an apparent estimate beyond the no information threshold.

Alternatively, optimism correction estimates optimism as the average difference between performance of bootstrap models as evaluated in a) the original full data set and b) within the bootstrap sample. In case of overfitting, the discrepancy between the two will increase. The apparent estimate is subsequently corrected for this bootstrap estimate of optimism. 

Obtaining either the 0.632+ or optimism corrected estimates for cben-$\hat{\delta}$ and cben-$\hat{y}^0$ is straightforward. One subtlety is that, in case of unequal group sizes (treated vs control), the average over 1000 repeated analyses of subsamples of the larger arm was taken to accommodate for 1:1 matching. For the model-based estimates, a choice with respect to the estimation of $\hat{P}_{\textnormal{benefit}, k, l}$ has to be made with respect to out-of-sample evaluation. To avoid bias, the out-of-sample evaluation of $\hat{P}_{\textnormal{benefit}, k, l}$ for the 0.632+ estimate was based on a model for $\hat{g}_0$ and $\hat{g}_1$ fitted in the out-of-sample cases (with the same specification as the model under evaluation). For the optimism correction, $\hat{P}_{\textnormal{benefit}, k, l}$ for original full data was based on the ITE model as developed in the full development data. That is, the 0.632+ model-based c-statistic estimates were obtained from (1) out-of-sample predictions $\hat{\delta}(\bm{x}_{i \in oos})$ from bootstrap ITE models and (2) $\hat{P}_{\textnormal{benefit}, k, l}$ based on an out-of-sample model. Optimism corrected model-based c-statistic estimates were obtained from (1) predictions $\hat{\delta}(\bm{x}_{i})$ from bootstrap models and $\hat{P}_{\textnormal{benefit}, k, l}$ based on the development model.

\textit{Calibration} \newline
Bootstrap evaluation of the calibration parameters was also performed. A 0.632+ estimate was derived for the slope estimates in analogy to the derivation for c-statistics, but using $\gamma=0$ for the value that slope $\beta_1$ takes for an uninformative model. Out-of-sample estimates of $\hat{g}_0(\cdot)$ and $\hat{g}_1(\cdot)$ were based on just the out-of-sample cases to serve as an offset in the calibration model. A 0.632+ estimate for the calibration intercept parameter is not readily available since a $\gamma$ value for a non-informative intercept cannot be defined. Optimism corrected bootstrap estimates were obtained for both intercepts and slopes. Estimates of $\hat{g}_0(\cdot)$ and $\hat{g}_1(\cdot)$ for the in-sample cases were based on the bootstrap model; $\hat{g}_0(\cdot)$ and $\hat{g}_1(\cdot)$ for the full data set for optimism correction were based on the full sample ITE model. 

\subsection{External validation}
\textit{Discrimination} \newline
External validation was performed in both V1 (DGM-1) and V2 (DGM-2). ITE predictions can be evaluated directly using cben-$\hat{\delta}$ and cben$_opte$. For cben-$\hat{y}^0$, external estimates of control outcome risk $\hat{g}_0(\cdot)$ are required for matching purposes, and were obtained by refitting of the relevant parts of model \eqref{eq:ITEmodel} in the control arm (\textit{i.e.}, omitting parameters relating to $a$ which equal 0 for controls) in the external data. Note that fitting a new model will in general remove bias, but may have a high cost in terms of variance if the external data set is small. For the model-based c-for-benefit (mbcb), the accuracy of $\hat{P}_{\textnormal{benefit}, k, l}$, and hence the underlying $\hat{g}_0(\cdot)$ and $\hat{g}_1(\cdot)$, is paramount. In general, these estimates should be based on data independent from the ITE model under evaluation. Thereto, model \eqref{eq:ITEmodel} was refitted in the external data to obtain the required independent estimates $\hat{P}_{\textnormal{benefit}, k, l}$. 

\textit{Calibration} \newline
Direct calibration assessment in external data exactly followed the lines of apparent calibration assessment with all predictions (both $\hat{\delta}_{lp}(\bm{x}_j)$ and $\hat{g}_{lp,0}(\bm{x}_j)$ and $\hat{g}_{lp,1}(\bm{x}_i)$) based on the ITE model as derived in the development data and applied in V1 and V2. For external validation, the required independent estimates $\hat{g}_{lp,0}(\bm{x}_j)$ and $\hat{g}_{lp,1}(\bm{x}_i)$ were obtained from the validation samples based on a refit of model \eqref{eq:ITEmodel}.

\section[Calibration of direct individualized treatment effect predictions]{Calibration of direct individualized treatment effect \\predictions} \label{app:tian}

When prediction $\hat{\delta}(\bm{x})$ are obtained without accompanying potential outcome predictions, it is still possible to estimate a calibration intercept and slope in 1:1 randomized trial settings based on work by Tian et al. \cite{tian_simple_2014}. 

In their work, focus is on treatment effect estimation by means of transformed covariate models. Using their notation, let $T$ be -1 for controls and 1 for the treated, let $\bm{Z}$ be a q-dimensional covariate vector, Y the outcome, and $\bm{W}(\cdot)$ a p-dimensional function of the covariates, with shorthand $\bm{W}_i$ for subject $i$. 

For the continuous case, a simple multivariate regression with centered treatment effect interactions is 
\begin{equation*}
  Y = \beta_0' \mathbf{W}(\mathbf{Z}) + \gamma_0'\mathbf{W}(\mathbf{Z}) \cdot T/2 + \epsilon
\end{equation*}

Note that an intercept is \textit{included} in $\mathbf{W}(\mathbf{Z})$. Under this model, treatment effects conditional on covariate vector $\mathbf{z}$ are $\Delta(\mathbf{z}) = E(Y^{(1)} - Y^{(-1)}|\bm{Z}=\bm{z}) = \gamma_0'\mathbf{W}(\mathbf{z})$. Working model 
\begin{equation*}
  Y = \gamma_0'\bm{W}^* + \epsilon, 
\end{equation*}
with $\bm{W}^*_i = \bm{W}_i \cdot T/2$ allows estimation of $\gamma_0$, and, in turn, $\hat{\gamma}_0 \mathbf{W}(\mathbf{z})$ provides estimates $\hat{\Delta}(\bm{z})$ \cite{tian_simple_2014}. Now for calibration assessment of the estimates $\hat{\Delta}(\bm{z})$ in new (1:1 randomized) data, the observed outcomes can be regressed on $\beta_0 T + \beta_1 \hat{\Delta}(\bm{z}) \cdot T/2$ and the resulting intercept and slope provide a calibration intercept and slope for the predicted treatment effects. 

For the binary case, with 
\begin{equation*}
 P(Y=1 | \bm{Z},T) = \frac{\textnormal{exp}(\gamma_0'\bm{W}^*)}{1+\textnormal{exp}(\gamma_0'\bm{W}^*)}   
\end{equation*}
a logistic model with linear predictor $\gamma_0'\bm{W}^*$ can provide the estimate of  $\gamma_0$. Under quite general assumptions \cite{tian_simple_2014}, 
\begin{equation*}
 \hat{\Delta}(\bm{z}) = \frac{\textnormal{exp}(\hat{\gamma_0}'\bm{W}(\bm{z}) / 2) - 1}{\textnormal{exp}(\hat{\gamma_0}'\bm{W}(\bm{z}) / 2) + 1}
\end{equation*}
Thus, when given only $\hat{\Delta}(\bm{z})$, the inverse transformation 
\begin{equation*}
 2 \textnormal{ln}(-\frac{\hat{\Delta}(\bm{z})+1}{\hat{\Delta}(\bm{z})-1})
\end{equation*}
provides $\hat{\gamma_0}'\bm{W}(\bm{z})$ which is on the linear predictor scale for a logistic model. Hence, a logistic model
\begin{equation}
    \textnormal{logit}(P(Y=1)) = \gamma_0 T + \gamma_1 \cdot 2 \textnormal{ln} \left( -\frac{\hat{\Delta}(\bm{z})+1}{\hat{\Delta}(\bm{z})-1}\right) \cdot T/2 
\end{equation}
provides an estimated calibration intercept $\hat{\gamma_0}$ and slope $\hat{\gamma_1}$. Note that the estimand here is not the same as in the main text due to the lacking offsets and different transform of the treatment effect. Nonetheless, for an ITE model just providing estimated individualized treatment effects, the estimates $\hat{\gamma_0}$ and slope $\hat{\gamma_1}$ have the typical calibration interpretation with the intercept equal to zero and the slope equal to 1 for perfectly calibration predictions. Deviations of the slope and intercept also have a similar interpretation, but now corresponding to the new transformation.  

\textbf{Simulation}
A short simulation study was performed to assess the benefit of re-calibration based on (i) the method of calibration assessment in the main text, and (ii) using the method described above. Note that a direct comparison of the resulting estimates $\hat{\beta_0}$, $\hat{\beta_1}$ and $\hat{\gamma_0}$, $\hat{\gamma_1}$ is not meaningful since they have different estimands. 

For ease of reading, DGM-1 from the main text was used to generate all of the data. Also, the same ITE model was used (\textit{i.e.} equation \eqref{eq:ITEmodel}). Sample size of the development sets was varied from small to large (100, 250, 500, 750, 1000). Predictions $\hat{\bm{\delta}}$ from the fitted ITE model were assessed in independent validation data of size $n=1000$ in term of parameters $\hat{\beta_0}$, $\hat{\beta_1}$, $\hat{\gamma_0}$ and $\hat{\gamma_1}$. These were used to update to linearly update $\hat{\bm{\delta}}$ on the appropriate scale (\textit{i.e.} using the relevant transformation for each calibration method). Subsequently, the updated predictions $\hat{\bm{\delta}}^*$ were evaluated in a third set of independent data of size $n=1000$ in terms of root mean squared error of $\hat{\bm{\delta}}^*$ with respect to the true $\bm{\delta}$. Table \ref{app:tabCalTian} shows that the rmse of the unadjusted error decreased with increasing sample size for model development. Updated ITE prediction based on the method in the main text (\textit{i.e.} based on $\hat{\beta_0}$ and $\hat{\beta_1}$) decreased rmse of the ITE predictions for small samples, and did so slightly better than updating based on $\hat{\gamma_0}$ and $\hat{\gamma_1}$. We hypothesize that this is due to the simple functional form of the true outcome risks, that provide a favourable effect of using the estimates $\hat{g}_0(\cdot)$ and $\hat{g}_1(\cdot)$. However, we hypothesize that reliance of the main method on $\hat{g}_0(\cdot)$ and $\hat{g}_1(\cdot)$ can also harm the estimator, in particular when the outcome risk has a difficult form or the model is misspecified. In such cases, calibration based just on estimates $\hat{\bm{\delta}}$ as presented here would be preferable. In practice, it is hard to know in which of these two settings you are. 

Extension to observational data can be envisioned based on a modification of the work by Tian et al. for observational data \cite{chen_general_2017}. Also, recent conference proceedings by Xu et al. provides quite general procedures for calibration of direct individualized treatment effect predictions \cite{xu_calibration_2022}. As opposed to the calibration methods typically used in epidemiological studies, they propose a nonparametric estimator that can be applied to continuous, binary and survival settings. This seems especially relevant when there is treatment selection and when the model under evaluation is very flexible and does not assume a specific generalized linear model structure. Future work may examine the possible gains of the parametric assumptions when valid, and whether this is offset by the risk of bias they always convey.  

\begin{table}[H] 
\centering
\begin{tabular}{lrrrrr} 
  Sample size & 100 & 250 & 500 & 750 & 1000 \\
  \hline   
\textbf{External} && \\
Unadjusted & 0.134 & 0.079 & 0.058 & 0.046 & 0.040 \\ 
Updated based on $\hat{\beta_0}$, $\hat{\beta_1}$ & 0.068 & 0.053 & 0.047 & 0.044 & 0.042  \\ 
Updated based on $\hat{\gamma_0}$, $\hat{\gamma_1}$ & 0.079 & 0.059 & 0.052 & 0.047 & 0.044 \\ 
   \hline
\end{tabular}
\caption{Simulation results on root mean squared error of $\hat{\bm{\delta}}^*$ with respect to the true $\bm{\delta}$. The median over n=500 simulations is shown.} \label{app:tabCalTian}
\end{table}

\section{Additional simulation study results}

\subsection{Discrimination} \label{app:suppl_discr_figures}

\begin{figure}[!htb]
\centering
\includegraphics[width=1\linewidth]{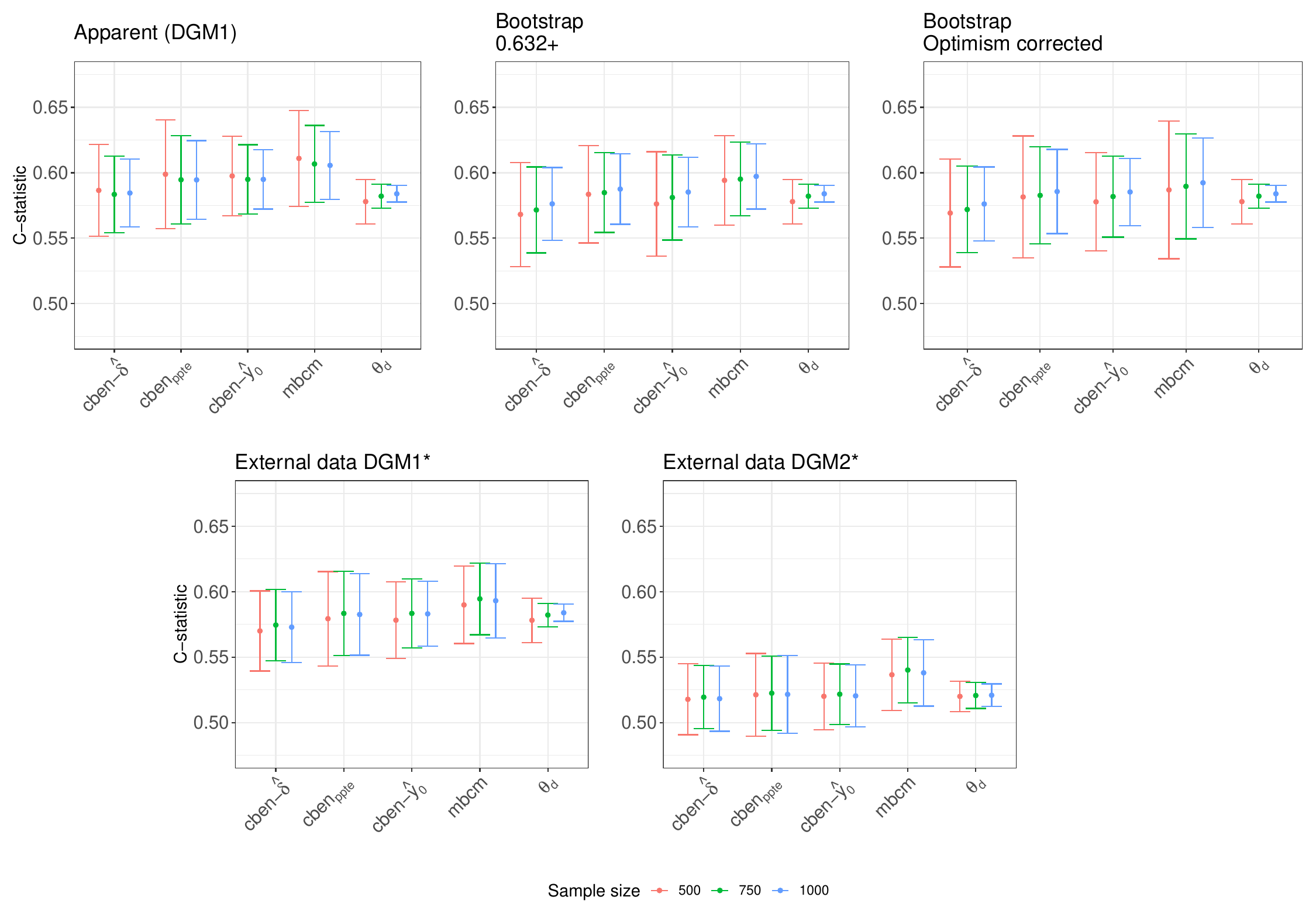}
\caption{Supplementary figures showing the mean $\pm$ 1 SD for the estimate c-statistics and estimand across simulations.}
\label{fig:simResults_discr_suppl}
\end{figure}

\subsection{Calibration} \label{app:suppl_cal_figures}

\begin{figure}[!htb]
\centering
\includegraphics[width=1\linewidth]{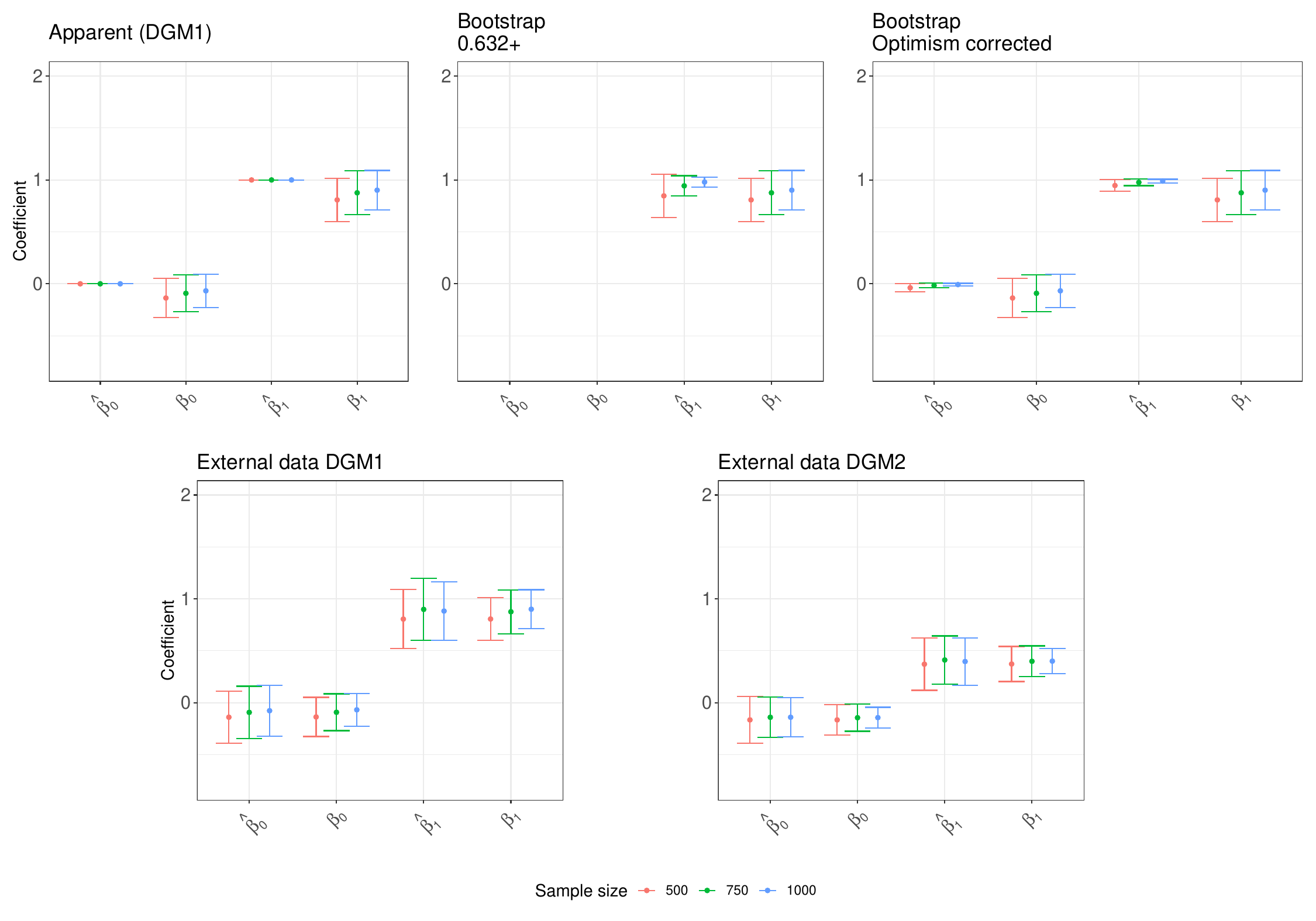}
\caption{Supplementary figures showing the 10\% trimmed mean $\pm$ 1 SD for the estimated calibration intercepts and slopes and the corresponding estimands across simulations.}
\label{fig:simResults_cal_suppl}
\end{figure}


\end{document}